\documentclass[a4paper]{article}

\usepackage{textcomp}
\usepackage{authblk}
\usepackage{hyperref}
\usepackage{cite}
\usepackage{tikz}
\usepackage{mathtools}
\usepackage{dsfont}
\usepackage{tikz}
\usepackage{graphicx}
\usepackage{dcolumn}
\usepackage{bm}

\usepackage{pifont}
\usepackage{amssymb}
\usepackage{amsmath}
\usepackage{amsthm,amsfonts}

\usepackage{hyperref}
\usepackage{epstopdf}
\usepackage{caption}
\usepackage{subcaption}
\usepackage{float}
\usepackage{enumitem}
\usepackage{calc}
\usepackage{caption}
\usepackage{subcaption}

\RequirePackage{graphicx}

\usepackage[margin=1in]{geometry}

\theoremstyle{plain}

\def\>>{\rangle}
\def\<<{\langle}
\def\tr{{\rm tr}}

\def\bfC{{\bf C}}
\def\bfJ{{\bf J}}
\def\bfH{{\bf H}}
\def\bfQ{{\bf Q}}
\def\bfP{{\bf P}}
\def\bfS{{\bf S}}

\def\bfrho{\boldsymbol \rho}
\def\bflambda{\boldsymbol \lambda}
\def\bfsigma{\boldsymbol \sigma}

\begin{document}

\title{An integrable approach to macroscopic fluctuation theory for the multispecies SSEP}

\date{} 

\author[1]{Luigi Cantini\thanks{\href{mailto:luigi.cantini@cyu.fr}{luigi.cantini@cyu.fr}}}
\affil[1]{Laboratoire de Physique Th\'eorique et Mod\'elisation, CNRS UMR 8089, CY Cergy Paris Universit\'e, 95302 Cergy-Pontoise Cedex, France}

\maketitle

\begin{abstract}
We study the macroscopic fluctuation theory (MFT) of a multispecies generalization of the symmetric simple exclusion process (mSSEP) on the infinite line, in which particles of $N+1$ species -- including, possibly, a vacancy species -- exchange positions at unit rate. Working with the full, redundant set of coarse-grained densities $\bfrho=\{\rho_0,\dots,\rho_N\}$ keeps the relabelling symmetry of the model manifest throughout. We first extend the argument of Derrida and Gerschenfeld to the multispecies setting, showing that the 
cumulant generating function of the multispecies current between two 
regions of an arbitrary graph depends on the boundary densities $
\bfrho_L,\bfrho_R$ and on the fugacities $\bflambda$ only through a 
single scalar variable $\omega$. We then formulate the MFT saddle-
point equations for the mSSEP on the infinite line and show that they 
define an integrable system: they are of Landau--Lifshitz type, and a 
Zakharov--Takhtajan gauge transformation recasts them in AKNS form.
Solving the resulting linear scattering problem by the inverse 
scattering method, we recover the cumulant generating function 
$F(\omega)$ for the multispecies current, as well as the initial and 
final density profiles conditioned on a prescribed current 
fluctuation. In particular, we show that $F(\omega)$ coincides with 
the function obtained by Derrida and Gerschenfeld for the single-
species SSEP, now derived for an arbitrary number of species directly 
from the integrable structure of the multispecies MFT equations.
\end{abstract}

\section{Introduction}

Interacting particle systems in one dimension provide an ideal testing ground for methods in non-equilibrium statistical mechanics
\cite{katz1984nonequilibrium,spohn2012large,schutz2001exactly,derrida2007non}.
At the macroscopic scale such systems typically display hydrodynamic behavior \cite{kipnis2013scaling}, but characterizing their \emph{fluctuations} away from equilibrium requires considerably more refined tools.

Two complementary strategies have proven especially fruitful. The first exploits exact solvability of the underlying microscopic dynamics --- through the Bethe ansatz, matrix-product representations, or integrable probability --- and has produced a detailed picture of current and density fluctuations in the non-equilibrium steady state, both for finite systems on periodic geometries and for finite strips coupled to boundary reservoirs \cite{derrida2007non}. Exact results are also available away from the steady state, for genuinely non-stationary initial conditions \cite{chen2018exact,chen2022limiting}. The second strategy abandons microscopic exact solvability in favor of generality: macroscopic fluctuation theory (MFT) \cite{bertini2001fluctuations,bertini2002macroscopic,bertini2015macroscopic} provides a mesoscopic, large-deviation description that applies, in principle, to entire universality classes of diffusive systems, expressed purely in terms of a diffusion matrix and a mobility matrix.
In recent years MFT has been used to obtain exact results for diffusive systems well beyond what direct microscopic computations allow \cite{krapivsky2012fluctuations,krapivsky2014large}. 

Current and density fluctuations in non-equilibrium steady states have been studied extensively for systems on a finite interval coupled to two reservoirs, as well as on a ring, often building on exact solutions of microscopic models such as the weakly asymmetric and asymmetric simple exclusion processes (WASEP and ASEP). 

Another setting is that of an \emph{infinite} system, with either stationary or non-stationary initial data. There the natural observables are the time-integrated current --- the total particle flux through a fixed point (say, the origin) up to time $t$ --- and the position of a tagged (tracer) particle at time $t$. The goal is to characterize the fluctuations of these quantities through their cumulants, or more completely through the cumulant generating function (CGF) and the associated large deviation rate function \cite{derrida2007non,derrida2009current,poncet2021generalised,grabsch2022exact}. 

Solving the MFT saddle-point equations that determine these quantities is generally difficult; quite surprisingly, the saddle-point equations of MFT for several one 
component systems like SSEP or KMP model turn out to be classically 
integrable, and integrability techniques --- most 
notably the inverse scattering method --- have been successfully 
employed to solve them exactly \cite{krajenbrink2021inverse,krajenbrink2022inverse,
bettelheim2022inverse,bettelheim2022full,mallick2022exact,mallick2024exact,
krajenbrink2026integrability,bettelheim2024complete}.

While MFT in principle accommodates diffusive \emph{multicomponent} systems on the same footing as single-species ones, this generalization remains comparatively little explored. To our knowledge the cases studied in this framework are limited to the ABC model \cite{bodineau2011phase,bodineau2008long}, together with related applications to active matter \cite{agranov2021exact,agranov2023macroscopic}.

In the present paper we study a multispecies generalization of the symmetric simple exclusion process (SSEP), which we refer to as the mSSEP. The model is defined on an arbitrary graph: each vertex is occupied by a particle of one of $N+1$ species (including, possibly, a vacancy species), and particles of different species sitting on neighboring vertices exchange positions in continuous time with rate $1$. On a finite strip coupled to reservoirs, this model was considered at the microscopic level by Vanicat \cite{vanicat2017exact}, who obtained the exact matrix-product steady state and, from it, the current, the density correlations, and the large deviation functional of the density profile, matching the MFT predictions.
Our approach extends to the multispecies setting the integrability-based analysis of \cite{mallick2022exact}, developed there for the single-species SSEP on the infinite line, and we believe that it sheds light on the origin of the integrable structure underlying the single-species MFT saddle-point equations themselves.

The dynamics of the mSSEP is invariant under relabelling of the species, and a central feature of our analysis is to preserve this invariance manifestly by working throughout with the full, redundant set of coarse-grained densities $\bfrho=\{\rho_0,\rho_1,\dots,\rho_N\}$, subject only to the exclusion constraint $\sum_\alpha \rho_\alpha = 1$. In this way the integrability of the MFT saddle point equations 
emerges quite naturally.

Denoting by $C_\alpha(t)$ the total charge of species $\alpha$ transferred from left to right up to time $t$, the long-time behavior of the moment generating function is expected to take the form
\begin{equation}
\left\langle
\exp\!\left[
\sum_{\alpha=0}^{N}\lambda_\alpha C_\alpha(t)
\right]
\right\rangle \sim e^{\sqrt{t}\,F(\bfrho_L,\bfrho_R,\bflambda)}.
\end{equation}
One of our main results is that the large deviation function $F(\bfrho_L,\bfrho_R,\bflambda)$ depends on the initial densities $\bfrho_L,\bfrho_R$ and on the fugacities $\bflambda$ only through the single combination
\begin{equation}\label{omega-dep}
\omega = \left(\sum_{k=0}^{N}\rho_{L,k}e^{\lambda_k}\right)
\left(\sum_{k=0}^{N}\rho_{R,k}e^{-\lambda_k}\right)-1,
\end{equation}
so that
\begin{equation}
F(\bfrho_L,\bfrho_R,\bflambda) = F(\omega),
\end{equation}
thereby extending to the multispecies setting the result obtained for the single-species SSEP by Derrida and Gerschenfeld \cite{derrida2009current}.

The paper is organized as follows. In Section~\ref{DG-section} we extend the Derrida--Gerschenfeld argument \cite{derrida2009current}, originally devised for the single-species SSEP, to the multispecies case, showing that the generating function of the total flux $\bfC(T)$ between two regions of an arbitrary graph depends on the densities $\bfrho_L$, $\bfrho_R$ and the fugacities $\bflambda$ only through the parameter $\omega$ defined in~\eqref{omega-dep}. In Section~\ref{MFT-SSEP} we formulate the MFT for the multispecies SSEP on the infinite line and derive the corresponding saddle-point equations. In Section~\ref{lax-section} we show that these equations are integrable by exhibiting an associated Lax pair of Landau--Lifshitz type, and we recast the problem, via a Zakharov--Takhtajan gauge transformation, in AKNS form so as to incorporate the relevant boundary conditions. In Section~\ref{linearization-sect} we solve the resulting equations using the inverse scattering method; this not only reproduces the current cumulant generating function but also yields the initial and final density profiles conditioned on a prescribed current fluctuation. Section~\ref{concl-sect} contains our conclusions and an outlook on possible extensions. The appendices collect, in Appendix~\ref{simple-ex}, a fully worked out example illustrating the general result of Section~\ref{DG-section}, and, in the remaining appendices, technical material that complements the derivations of the main text.

\section{$\omega$ dependence of the current moment generating function}
\label{DG-section}

Let us consider the multispecies symmetric simple exclusion process (SSEP) on a general graph $G$. We partition the vertex set of $G$ into two disjoint subsets $L$ and $R$, with $R = L^c$. 
Assume that at time $t=0$ the system is initialized in a completely factorized (product) measure. More precisely, for each site $x \in L$, the site is independently occupied by a particle of species $\alpha$ with probability $\rho_{L,\alpha}$, while for each site $x \in R$, the site is independently occupied by a particle of species $\alpha$ with probability $\rho_{R,\alpha}$. 
We collect these parameters into the vectors 
$\bfrho_L = \{\rho_{L,0}, \rho_{L,1}, \dots, \rho_{L,N}\}$ 
and 
$\bfrho_R = \{\rho_{R,0}, \rho_{R,1}, \dots, \rho_{R,N}\}$, 
which satisfy the normalization condition
\begin{equation}\label{constr-initial}
\sum_{\alpha=0}^N \rho_{L,\alpha} = \sum_{\alpha=0}^N \rho_{R,\alpha} = 1.
\end{equation}

\medskip
\noindent
We define the \emph{net inflow} of particles of species $\alpha$ into the region $R$ during the time interval $[0,t]$ as
\begin{equation}
C_\alpha(t)
=
\sum_{x \in R} \bigl( \eta_\alpha(x,t) - \eta_\alpha(x,0) \bigr),
\end{equation}
where $\eta_\alpha(x,t)$ denotes the occupation variable of species $\alpha$ at site $x$ and time $t$, i.e.\ $\eta_\alpha(x,t)=1$ if the site is occupied by a particle of species $\alpha$ at time $t$, and $\eta_\alpha(x,t)=0$ otherwise. 
Since each site is occupied by exactly one particle, we have
\[
\sum_{\alpha=0}^N \eta_\alpha(x,t) = 1,
\]
which implies the conservation law
\begin{equation}\label{zero-total-flow}
\sum_{\alpha=0}^N C_\alpha(t) = 0.
\end{equation}
%
%
Our goal in this section is to show, following and generalizing the argument of Derrida and Gerschenfeld \cite{derrida2009current}, that the moment generating function of the currents,
\begin{equation}
f(\bflambda,\bfrho_L,\bfrho_R;t)
=
\left\langle 
\exp\!\left(
\sum_{\alpha=0}^{N}\lambda_\alpha C_\alpha(t)
\right)
\right\rangle,
\qquad 
\bflambda=\{\lambda_0,\lambda_1,\dots,\lambda_N\},
\end{equation}
depends on the initial densities $\bfrho_L$, $\bfrho_R$ and the fugacities
$\bflambda$ only through a single scalar parameter,
\begin{equation}
\omega
=
\left(\sum_{\alpha=0}^{N}\rho_{L,\alpha}e^{\lambda_\alpha}\right)
\left(\sum_{\alpha=0}^{N}\rho_{R,\alpha}e^{-\lambda_\alpha}\right)-1.
\end{equation}
In particular, this implies that computing $f$ in the single-species case ($N=1$) suffices to recover the result for arbitrary $N$.

\medskip
\noindent
We write
\[
f(\bflambda,\bfrho_L,\bfrho_R;t)
=
f(\lambda_0,\lambda_1,\dots,\lambda_N;
\overline\rho_0,\overline\rho_1,\dots,\overline\rho_N;t),
\]
where we introduce the shorthand
$
\overline\rho_\alpha=(\rho_{L,\alpha},\rho_{R,\alpha})
$.

\medskip
\noindent
\textbf{Permutation symmetry.}
Since the dynamics treats all species symmetrically, the generating function is invariant under permutations of the species labels. For any permutation $\sigma \in \mathcal{S}_{N+1}$,
\begin{equation}\label{perm-symm}
f(\lambda_0,\dots,\lambda_N;
\overline\rho_0,\dots,\overline\rho_N;t)
=
f(\lambda_{\sigma(0)},\dots,\lambda_{\sigma(N)};
\overline\rho_{\sigma(0)},\dots,\overline\rho_{\sigma(N)};t).
\end{equation}

\medskip
\noindent
Because of the normalization condition \eqref{constr-initial}, the variables $\overline\rho_1,\dots,\overline\rho_N$ can be taken as independent. Moreover, the constraint \eqref{zero-total-flow} implies that $f$ is invariant under a uniform shift of all counting fields,
\[
\lambda_\alpha \;\longrightarrow\; \lambda_\alpha + \varepsilon.
\]
Hence $f$ depends only on $N$ independent combinations of the $\lambda_\alpha$. A convenient choice is
\[
\nu_i = \lambda_i - \lambda_0, \qquad i=1,\dots,N.
\]
We may therefore rewrite
\begin{equation}
f(\lambda_0,\dots,\lambda_N;
\overline\rho_0,\dots,\overline\rho_N;t)
=
g(\nu_1,\dots,\nu_N;
\overline\rho_1,\dots,\overline\rho_N;t).
\end{equation}
The permutation symmetry \eqref{perm-symm} induces a corresponding symmetry for $g$, together with an additional relation obtained by exchanging species $0$ and $N$:
\begin{equation}\label{inv-g}
g(\nu_1,\dots,\nu_N;
\overline\rho_1,\dots,\overline\rho_N;t)
=
g(\nu_1-\nu_N,\dots,-\nu_N;
\overline\rho_1,\dots,1-\sum_{\alpha=1}^{N}\overline\rho_\alpha;t).
\end{equation}

\medskip
\noindent
\textbf{Derrida--Gerschenfeld expansion.}
Following \cite{derrida2009current}, we expand the generating function as
\begin{equation}
f(\bflambda;\overline{\bfrho};t)
=
\sum_{\substack{|\underline{p_L}|=N_L\\|\underline{p_R}|=N_R}}
\prod_{\alpha=0}^{N}
\rho_{L,\alpha}^{p_{L,\alpha}}
\rho_{R,\alpha}^{p_{R,\alpha}}
\;
M_{\underline{p_L},\underline{p_R}}
(e^{\lambda_0},\dots,e^{\lambda_N};t),
\end{equation}
where $N_L$ and $N_R$ denote the number of sites in $L$ and $R$, and
\begin{equation*}
\underline{p_L}=(p_{L,0},\dots,p_{L,N}), \qquad
\underline{p_R}=(p_{R,0},\dots,p_{R,N}),
\end{equation*}
with $|\underline{p_L}|=\sum_\alpha p_{L,\alpha}$ and $|\underline{p_R}|=\sum_\alpha p_{R,\alpha}$.The function
$
M_{\underline{p_L},\underline{p_R}}(x_0,\dots,x_N;t)
$
is a homogeneous Laurent polynomial of total degree zero in the variables $x_\alpha$, with degree in $x_\alpha$ lying in the interval
$
[-p_{R,\alpha},\,p_{L,\alpha}].
$
This reflects the constraint that at most $p_{L,\alpha}$ particles of species $\alpha$ can enter $R$, and at most $p_{R,\alpha}$ can leave it.

\medskip
\noindent
Eliminating species $0$, we obtain
\begin{equation}
g(\underline{\nu};\underline{\bfrho};t)
=
\sum_{\substack{|\underline{p_L}|\le N_L\\|\underline{p_R}|\le N_R}}
\prod_{\alpha=1}^{N}
\rho_{L,\alpha}^{p_{L,\alpha}}
\rho_{R,\alpha}^{p_{R,\alpha}}
\,
S_{\underline{p_L},\underline{p_R}}
(e^{\nu_1},\dots,e^{\nu_N};t),
\end{equation}
where $S_{\underline{p_L},\underline{p_R}}$ is a Laurent polynomial with degree bounds
$[-p_{R,\alpha},\,p_{L,\alpha}]$ in each variable.
Using the same reasoning as in \cite{derrida2009current}, one finds that
$\langle \prod_{\alpha=1}^N C_\alpha^{n_\alpha}(t)\rangle$ is a polynomial of degree $n_\alpha$ in $\rho_{L,\alpha}$ and $\rho_{R,\alpha}$. Consistency then requires
\begin{equation}
S_{\underline{p_L},\underline{p_R}}
(e^{\nu_1},\dots,e^{\nu_N};t)
=
O\!\left(\prod_{\alpha=1}^N \nu_{\alpha}^{p_{L,\alpha}+p_{R,\alpha}}\right),
\end{equation}
which implies the factorized form
\begin{equation}
S_{\underline{p_L},\underline{p_R}}
=
s_{\underline{p_L},\underline{p_R}}(t)
\prod_{\alpha=1}^{N}
\left(e^{\nu_\alpha}-1\right)^{p_{L,\alpha}}
\left(e^{-\nu_\alpha}-1\right)^{p_{R,\alpha}}.
\end{equation}
Thus,
\begin{equation}
g(\underline{\nu};\underline{\bfrho};t)
=
\sum
s_{\underline{p_L},\underline{p_R}}(t)
\prod_{\alpha=1}^{N}
\left(\rho_{L,\alpha}(e^{\nu_\alpha}-1)\right)^{p_{L,\alpha}}
\left(\rho_{R,\alpha}(e^{-\nu_\alpha}-1)\right)^{p_{R,\alpha}}.
\end{equation}
Therefore $g$ depends on the parameters only through the combinations
\begin{equation}
x_\alpha=\rho_{L,\alpha}(e^{\nu_\alpha}-1),
\qquad
y_\alpha=\rho_{R,\alpha}(e^{-\nu_\alpha}-1),
\end{equation}
so that
\begin{equation}
g(\underline{\nu};\underline{\bfrho};t)
=
G(x_1,\dots,x_N;y_1,\dots,y_N;t).
\end{equation}

\medskip
\noindent
Using the symmetry \eqref{inv-g}, one shows that $G$ satisfies
\begin{equation}\label{inv_G}
G(x_1,\dots,x_N;y_1,\dots,y_N;t)
=
G(\tilde x_1,\dots,\tilde x_N;\tilde y_1,\dots,\tilde y_N;t),
\end{equation}
for 
\[
\tilde x_n=(e^{\nu_n}-1)\left(1-\sum_{i=1}^n\frac{x_i}{e^{-\nu_i}-1}  \right), \qquad\tilde x_k= \frac{e^{\nu_n-\nu_k}-1}{e^{-\nu_k}-1}x_k\quad 1\leq k \leq n-1,
\]
\[
\tilde y_n=(e^{-\nu_n}-1)\left(1-\sum_{i=1}^n\frac{y_i}{e^{\nu_i}-1}  \right), \qquad\tilde y_k= \frac{e^{\nu_k-\nu_n}-1}{e^{\nu_k}-1}y_k\quad 1\leq k \leq n-1.
\]
Since this holds for arbitrary $\nu_i$, we may set $\nu_i=\nu$, leading to a drastic simplification.
Choosing $e^{-\nu}=1+\sum_{k=1}^N y_k$, we finally obtain
\begin{equation}
G(x_1,\dots,x_N;y_1,\dots,y_N)
=
G\left(
0,
\left(1+\sum_{k=1}^{N}x_k\right)
\left(1+\sum_{k=1}^{N}y_k\right)-1;
0,0
\right).
\end{equation}
The generating function therefore depends only on the single scalar
\begin{equation}
\omega
=
\left(
\sum_{k=1}^{N}\rho_{L,k}e^{\nu_k}
+
1-\sum_{k=1}^{N}\rho_{L,k}
\right)
\left(
\sum_{k=1}^{N}\rho_{R,k}e^{-\nu_k}
+
1-\sum_{k=1}^{N}\rho_{R,k}
\right)-1,
\end{equation}
or equivalently,
\begin{equation}\label{omega-first}
\omega
=
\left(\sum_{k=0}^{N}\rho_{L,k}e^{\lambda_k}\right)
\left(\sum_{k=0}^{N}\rho_{R,k}e^{-\lambda_k}\right)-1.
\end{equation}
As a further consistency check, let us consider the case $N=1$. Setting $\lambda_1=\lambda$, $\lambda_0=0$, and writing
$\rho_1=\rho$, $\rho_0=1-\rho$, we recover the standard SSEP expression
\begin{equation}
\omega
=
(e^{\lambda}-1)\rho_L(1-\rho_R)
+
(e^{-\lambda}-1)\rho_R(1-\rho_L).
\end{equation}

\medskip
\noindent
Combining our result with the long-time asymptotics, we obtain
\begin{equation}
\left\langle 
\exp\!\left[
\sum_{\alpha=0}^{N}\lambda_\alpha C_\alpha(t)
\right]
\right\rangle 
\sim 
e^{\sqrt{t}\,F(\omega)}.
\end{equation}
Given the series expansion of $F(\omega)$ for small $\omega$,
\begin{equation}
F(\omega)= \sum_{n=1}^{\infty} f_n \frac{\omega^n}{n!},
\end{equation}
one can systematically compute the cumulants of the transferred particles. In particular, the first two cumulants read
\begin{gather}
\lim_{t\to \infty} t^{-\frac{1}{2}}\langle C_\alpha(t) \rangle_c 
= f_1 \,\Delta \rho_\alpha,\\
\lim_{t\to \infty} t^{-\frac{1}{2}}\langle C_\alpha(t)C_\beta(t) \rangle_c 
= f_2 \,\Delta \rho_\alpha \Delta \rho_\beta
+ f_1 \left(
\delta_{\alpha,\beta}(\rho_{L,\alpha}+\rho_{R,\alpha})
- \rho_{L,\alpha}\rho_{R,\beta}
- \rho_{L,\beta}\rho_{R,\alpha}
\right),
\end{gather}
where $\Delta \rho_\alpha = \rho_{L,\alpha} - \rho_{R,\alpha}$.

\medskip
\noindent
As discussed above, the function $F(\omega)$ can be determined by restricting to the single-species case. An important example is the SSEP on the infinite line, with step initial condition: each site to the left of the origin ($x \leq 0$) is occupied with probability $\bfrho_L$, while each site to the right ($x > 0$) is occupied with probability $\bfrho_R$. 
In this setting, $F(\omega)$ was computed in \cite{derrida2009current} using the Bethe Ansatz for the single-species SSEP, yielding
\begin{equation}
F(\omega)= \frac{1}{\sqrt{\pi}}\sum_{n\geq 1} \frac{(-1)^{n+1}}{n^{3/2}} \,\omega^n
= \frac{1}{\pi} \int_{-\infty}^{+\infty} dk \,\log \bigl(1+\omega e^{-k^2}\bigr).
\end{equation}
This result was later recovered within the macroscopic fluctuation theory (MFT) framework for the single-species SSEP in \cite{mallick2022exact}. As we will show below, it can also be derived by applying MFT to the multispecies SSEP.

\section{MFT for multispecies  SSEP}\label{MFT-SSEP}

We now turn to the study of the multispecies SSEP on 
a one-dimensional infinite lattice within the 
framework of macroscopic fluctuation theory (MFT).  
This theory applies to diffusive multicomponent 
systems and predicts that the probability of observing 
a given macroscopic evolution of the coarse-grained 
density profile $\bfrho(x,t)=\{\rho_0(x,t),\dots,
\rho_N(x,t)\}$ and current $\bfJ(x,t)=\{J_0(x,t),
\dots,J_N(x,t)\}$ is given by 
\cite{bertini2001fluctuations,bertini2002macroscopic,bertini2015macroscopic}
\begin{equation}\label{proba-1}
\mathbb{P}[\bfrho,\bfJ]
\propto
\exp\left[
-
\int_0^T dt \int_{-\infty}^{+\infty} dx \;
(\bfJ-\mathbf{D}\,\partial_x \bfrho)^\top
\bfsigma(\bfrho)^{-1}
(\bfJ-\mathbf{D}\,\partial_x \bfrho)
\right],
\end{equation}
subject to the continuity equation
\begin{equation}\label{continuity}
\partial_t \bfrho(x,t) + \partial_x \bfJ(x,t) = 0.
\end{equation}
The only quantities that depend on the specific model are the diffusion matrix $\mathbf{D}$ and the mobility matrix $\bfsigma(\bfrho)$.

\medskip
\noindent
For the multispecies SSEP, it is convenient to adopt a redundant description involving the full set of densities $\bfrho=\{\rho_0,\rho_1,\dots,\rho_N\}$ and currents $\bfJ=\{J_0,J_1,\dots,J_N\}$. These fields satisfy the constraints that the total density is identically equal to $1$ and that the total current vanishes \cite{vanicat2017exact}:
\begin{equation}\label{occ-constr}
\sum_{\alpha=0}^N \rho_\alpha(x,t)=1,
\qquad
\sum_{\alpha=0}^N J_\alpha(x,t)=0.
\end{equation}
For this model, the diffusion matrix reduces to the identity,
$
\mathbf{D}=\mathbf{I}
$.
Moreover, by the fluctuation--dissipation relation, the inverse mobility matrix takes the form
\begin{equation}
\bfsigma(\bfrho)^{-1}_{\alpha,\beta}
=
\frac{1}{4}
\,\frac{\partial^2 f(\bfrho)}
{\partial \rho_\alpha \partial \rho_\beta},
\end{equation}
where the free-energy functional is purely entropic:
\begin{equation}
f(\bfrho)
=
\sum_{\alpha=0}^N
\rho_\alpha \log \rho_\alpha.
\end{equation}

\medskip
\noindent
Introducing the shorthand
$
\overline{J}_\alpha = J_\alpha + \partial_x \rho_\alpha,
$
the integrand in \eqref{proba-1} can be rewritten as
\begin{equation}
\begin{split}
\mathcal{L}[\bfrho,\bfJ]
&=
\frac{1}{4}
\sum_{\alpha,\beta=1}^N
\overline{J}_\alpha
\Bigg(
\frac{\partial^2 f}{\partial \rho_\alpha \partial \rho_\beta}
+
\frac{\partial^2 f}{\partial \rho_0 \partial \rho_\beta}
\frac{\partial \rho_0}{\partial \rho_\alpha}
+
\frac{\partial^2 f}{\partial \rho_\alpha \partial \rho_0}
\frac{\partial \rho_0}{\partial \rho_\beta}
+
\frac{\partial^2 f}{\partial \rho_0^2}
\frac{\partial \rho_0}{\partial \rho_\alpha}
\frac{\partial \rho_0}{\partial \rho_\beta}
\Bigg)
\overline{J}_\beta .
\end{split}
\end{equation}
Using the constraint
$
\rho_0 = 1 - \sum_{\alpha=1}^N \rho_\alpha,
\frac{\partial \rho_0}{\partial \rho_\alpha} = -1,
$
together with the identity
$
\sum_{\alpha=0}^N \overline{J}_\alpha = 0,
$
this expression simplifies to
\begin{equation}
\mathcal{L}[\bfrho,\bfJ]
=
\frac{1}{4}
\sum_{\alpha,\beta=0}^N
\overline{J}_\alpha
\frac{\partial^2 f}{\partial \rho_\alpha \partial \rho_\beta}
\overline{J}_\beta
=
\sum_{\alpha=0}^N
\frac{\left[J_\alpha(x,t)+\partial_x \rho_\alpha(x,t)\right]^2}
{4\rho_\alpha(x,t)}.
\end{equation}

\medskip
\noindent
In summary, the probability of observing a given space--time evolution of the density and current fields is given by \cite{vanicat2017exact}
\begin{equation}
\mathbb{P}[\bfrho,\bfJ]
\propto
\exp\left[
-
\int_0^T dt
\int_{-\infty}^{+\infty} dx \;
\mathcal{L}[\bfrho(x,t),\bfJ(x,t)]
\right],
\end{equation}
subject to the continuity equation \eqref{continuity} and the occupation constraint \eqref{occ-constr}.

\medskip
\noindent
In the following, we formulate the analysis entirely within this redundant description, retaining all species densities and currents. Besides preserving the full permutation symmetry between species, this representation also makes the connection with the standard single-species SSEP formulation more transparent.

\medskip
\noindent
The probability of observing a given density profile must be complemented by the probability of the initial configuration, determined by the underlying microscopic measure. 
If the measure on an interval of $\ell$ sites is a product Bernoulli measure with probability $\overline{\rho}_\alpha$ of finding a particle of species $\alpha$ at each site, then
\begin{equation}
P(\bfrho,\ell\vert \overline{\bfrho})
= \frac{\ell!}{(\rho_0 \ell)!(\rho_1 \ell)!\cdots(\rho_N \ell)!}
\prod_{\alpha=0}^N \overline{\rho}_\alpha^{\ell \rho_\alpha}
\approx 
\exp\left[
-\ell \sum_{\alpha=0}^N \rho_\alpha 
\log \left( \frac{\rho_\alpha}{\overline{\rho}_\alpha} \right)
\right],
\end{equation}
where we used Stirling’s approximation.

\medskip
\noindent
If the initial measure is still Bernoulli but with site-dependent probabilities $\overline{\bfrho}(x)$, we coarse-grain the system into boxes of size $\ell$ such that $1 \ll \ell \ll L$, where $L$ is the macroscopic scale. The probability of observing a coarse-grained profile $\bfrho(x)$ is then
\begin{equation}
P(\bfrho(x)\vert \overline{\bfrho}(x)) 
= \exp\left[ -F_0(\bfrho\vert \overline{\bfrho}) \right],
\end{equation}
with
\begin{equation}
F_0(\bfrho\vert \overline{\bfrho})
= \int_{-\infty}^{+\infty} dx 
\sum_{\alpha=0}^N \rho_\alpha(x)
\log \left( \frac{\rho_\alpha(x)}{\overline{\rho}_\alpha(x)} \right).
\end{equation}

\medskip
\noindent
Below we shall focus on step-like initial conditions: for $x<0$,
$\overline{\bfrho}(x,0)=\bfrho_L=(\rho_{L,0},\dots,\rho_{L,N})$, while for $x>0$,
$\overline{\bfrho}(x,0)=\bfrho_R=(\rho_{R,0},\dots,\rho_{R,N})$.
We are interested in the integrated current
\begin{equation}
\bfC(T)=\int_0^{+\infty} dx\, [\bfrho(x,T)-\bfrho(x,0)].
\end{equation}

\medskip
\noindent
Putting everything together, we consider the path integral representation of the cumulant generating function of the integrated current
\begin{equation}\label{path-int1}
\left\langle e^{\bflambda\cdot \bfC(T)}\right\rangle 
\propto \int \mathcal{D}\bfrho \,\mathcal{D}\bfJ \;
\delta\!\left(\sum_{\alpha=0}^N \rho_{\alpha}-1\right)
\delta\!\left(\sum_{\alpha=0}^N J_{\alpha}\right)
\delta(\partial_t \bfrho+ \partial_x \bfJ)\,
\exp \left(\mathcal{I}_{bulk}[\bfrho,\bfJ]+\mathcal{I}_{bdr}[\bfrho]\right),
\end{equation}
with action consisting of a bulk term
\begin{equation}
\mathcal{I}_{bulk}[\bfrho,\bfJ]
=   - \int_0^T dt \int_{-\infty}^{+\infty} dx \, \mathcal{L}[\bfrho,\bfJ].
\end{equation}
and a boundary term
\begin{equation}
\mathcal{I}_{bdr}[\bfrho]
=   \bflambda\cdot \bfC
- F_0[\bfrho(x,0)].
\end{equation}
Exponentiating the delta constraints by introducing the auxiliary fields $\bfH$, $k_\rho$, and $k_J$ renders the action quadratic in $\bfJ$, thereby allowing us to perform the Gaussian integration over $\bfJ$.  
The trade-off is the introduction of spurious degrees of freedom, which manifest as a gauge invariance in the resulting path integral.
Indeed, the continuity equation \eqref{continuity}, together with the constraint $\sum_\alpha \rho_\alpha = 1$, implies that
\[
\partial_x \left(\sum_{\alpha=0}^N J_\alpha \right) = 0,
\]
so that the total current is spatially uniform. Consequently, enforcing the vanishing of the total current via a local Lagrange multiplier introduces redundant local modes that do not correspond to physical fluctuations.
Proceeding with the Gaussian integration over $\bfJ$, we obtain
\begin{equation}
\left\langle e^{\bflambda\cdot \bfC(T)}\right\rangle
\propto
\int \mathcal{D}\bfrho \,\mathcal{D}\bfH \,\mathcal{D}k_\rho \,\mathcal{D}k_J \;
\exp\left[
\mathcal{I}_{\mathrm{bdr}}[\bfrho]
+
\overline{\mathcal{I}}_{\mathrm{bulk}}[\bfrho,\bfH,k_\rho,k_J]
\right],
\end{equation}
with
\begin{equation}
\overline{\mathcal{I}}_{\mathrm{bulk}}[\bfrho,\bfH,k_\rho,k_J]
\propto
- \int_0^T dt \int_{-\infty}^{+\infty} dx \;
\Big[
\bfH\cdot \partial_t \bfrho
+ \sum_\alpha (\partial_x H_\alpha - k_J)
\big(\partial_x \rho_\alpha - \rho_\alpha (\partial_x H_\alpha - k_J)\big)
+ k_\rho \big(\sum_\alpha \rho_\alpha - 1\big)
\Big].
\end{equation}
The dominant contribution is obtained from the stationary point of $\overline{\mathcal{I}}_{\mathrm{bulk}}$, which yields the equations of motion
\begin{subequations}
\begin{align}
\partial_t \rho_\alpha
&=
\partial_x \Big(\partial_x \rho_\alpha - 2\rho_\alpha (\partial_x H_\alpha - k_J)\Big),
\label{KPZ-like}
\\
\partial_t H_\alpha
&=
-\partial_x(\partial_x H_\alpha - k_J)
-(\partial_x H_\alpha - k_J)^2
+ k_\rho,
\\
\sum_{\alpha=0}^N \rho_\alpha &= 1,
\qquad
k_J = \sum_{\alpha=0}^N \rho_\alpha \partial_x H_\alpha.
\end{align}
\end{subequations}
with boundary conditions
\begin{subequations}
\begin{align}
H_\alpha(x,0)
&=
\theta(x)\lambda_\alpha
+ \log \rho_\alpha(x,0)
- \log \overline{\rho}_\alpha(x),
\\
H_\alpha(x,T)
&=
\theta(x)\lambda_\alpha.
\end{align}
\end{subequations}
As anticipated, the resulting equations of motion are invariant under the local gauge transformation
\begin{equation}\label{GT-1}
H_\alpha \to H_\alpha + \Lambda,
\qquad
\rho_\alpha \to \rho_\alpha,
\qquad
k_J \to k_J + \partial_x \Lambda,
\qquad
k_\rho \to k_\rho + \partial_t \Lambda,
\end{equation}
where $\Lambda$ is an arbitrary scalar field. The fields $k_J$ and $k_\rho$ therefore play the role of gauge potentials.

\medskip
\noindent
The structure of the action obtained above is that of a 
gauged nonlinear sigma model written in a non-standard  
set of variables. To make this structure more transparent, 
it is convenient to introduce new canonical fields $\bfQ$ 
and $\bfP$ defined by
\[
P_\alpha = e^{H_\alpha}, \qquad Q_\alpha = \rho_\alpha e^{-H_\alpha}.
\]
This change of variables trades the original density--response field pair $(\rho_\alpha,H_\alpha)$ for a pair $(Q_\alpha,P_\alpha)$ that plays the role of conjugate variables, and in which the gauge structure becomes explicit.
In terms of these variables, the bulk part of the action takes the compact form
\begin{equation}
\overline{\mathcal{I}}_{\mathrm{bulk}}[\bfQ,\bfP,k_J,k_\rho]
=
-\int_0^{T} dt \int_{-\infty}^{+\infty} dx \;
\Big(
- \bfQ \cdot \nabla^{-}_t \bfP
+ \nabla^{+}_x \bfQ \cdot \nabla^{-}_x \bfP
- k_\rho
\Big),
\end{equation}
where we have introduced the covariant derivatives
\[
\nabla^\pm_t = \partial_t \pm k_\rho,
\qquad
\nabla^\pm_x = \partial_x \pm k_J.
\]
These derivatives make the gauge structure manifest: the auxiliary fields $k_\rho$ and $k_J$ act as temporal and spatial gauge connections, respectively.
The corresponding Euler--Lagrange equations read
\begin{subequations}
\begin{align}
\label{time-Q}
\nabla^+_t \bfQ &= (\nabla^+_x)^2 \bfQ,\\
\label{time-P}
\nabla^-_t \bfP &= -(\nabla^-_x)^2 \bfP,\\
\label{cons-QP}
\bfQ \cdot \bfP &= 1,\qquad 
\bfQ \cdot \nabla^-_x\bfP - \bfP\cdot \nabla^+_x\bfQ = 0.
\end{align}
\end{subequations}
The first two equations describe covariant diffusion-like evolution for $\bfQ$ and $\bfP$, while the last line encodes the constraints inherited from the original occupation and current conservation laws.
Importantly, the system is invariant under the local gauge transformation
\[
\bfQ \to e^{-\Lambda} \bfQ,\qquad
\bfP \to e^{\Lambda} \bfP,\qquad
k_J \to k_J + \partial_x \Lambda,\qquad
k_\rho \to k_\rho + \partial_t \Lambda,
\]
confirming that $k_J$ and $k_\rho$ indeed play the role of gauge potentials.
Differentiating the constraint $\bfP\cdot\bfQ = 1$ with respect to $x$ yields
\begin{equation}
0 = \partial_x (\bfP\cdot \bfQ)
= \bfP\cdot\partial_x \bfQ + \bfQ\cdot \partial_x \bfP
= \bfP\cdot\nabla^+_x \bfQ + \bfQ\cdot \nabla^-_x \bfP,
\end{equation}
which, combined with \eqref{cons-QP}, implies the orthogonality relations
\begin{equation}\label{PdQ}
\bfP\cdot\nabla^+_x \bfQ = 0,
\qquad
\bfQ\cdot \nabla^-_x \bfP = 0.
\end{equation}
In terms of the variables $\bfQ$ and $\bfP$, the boundary conditions take a particularly simple form:
\begin{subequations}
\begin{align}\label{BCs00}
Q_\alpha(x,0) &= \overline{\rho}_\alpha(x)\, e^{-\theta(x)\lambda_\alpha},\\
\label{BCsT0}
P_\alpha(x,T) &= e^{\theta(x)\lambda_\alpha}.
\end{align}
\end{subequations}
In the case of step initial data,
\[
\overline{\bfrho}(x)= \bfrho_R\, \theta(x)+\bfrho_L\, \theta(-x),
\]
these boundary conditions simplify further to
\begin{subequations}
\begin{align}\label{BCs0}
\bfQ(x,0) &= \bfQ_R\, \theta(x)+\bfQ_L\,\theta(-x),\\
\label{BCsT}
\bfP(x,T) &= \bfP_R\, \theta(x)+\bfP_L\,\theta(-x),
\end{align}
\end{subequations}
with
\begin{equation}\label{boundary-values}
\begin{split}
\mathbf{Q}_L &= (\rho_{L,0},\dots,\rho_{L,N}), \qquad
\mathbf{Q}_R = (\rho_{R,0}e^{-\lambda_0},\dots,\rho_{R,N}e^{-\lambda_N}),\\
\mathbf{P}_L &= (1,\dots,1),\qquad
\mathbf{P}_R = (e^{\lambda_0},\dots,e^{\lambda_N}).
\end{split}
\end{equation}
These boundary conditions yield asymptotic conditions valid for all $t\in[0,T]$:
\begin{equation}\label{boundary-values2}
\begin{split}
\lim_{x\to -\infty}\mathbf{Q}(x,t) &= \mathbf{Q}_L, \qquad
\lim_{x\to +\infty}\mathbf{Q}(x,t) = \mathbf{Q}_R,\\
\lim_{x\to -\infty}\mathbf{P}(x,t) &= \mathbf{P}_L,\qquad
\lim_{x\to +\infty}\mathbf{P}(x,t) = \mathbf{P}_R.
\end{split}
\end{equation}

\subsection{Lax pair formulation and the Zakharov--Takhtajan gauge mapping}\label{lax-section}

The MFT equations \eqref{time-Q}--\eqref{cons-QP} can be recast in the form of the equations of motion of the Landau--Lifshitz model. To make this connection explicit, we introduce the gauge-invariant matrix field
\begin{equation}
\bfS = \bfQ \bfP^\top, \qquad \text{i.e.} \qquad \bfS_{\alpha\beta} = Q_\alpha P_\beta.
\end{equation}
By construction, the diagonal entries reproduce the physical densities,
\[
\bfS_{\alpha\alpha}(x,t) = \rho_\alpha(x,t).
\]
Moreover, using the constraint \eqref{cons-QP}, the matrix $\bfS$ is a rank-one projector:
\begin{equation}
\bfS^2 = \bfS, \qquad \mathrm{tr}(\bfS)=1.
\end{equation}
%
We now show that $\bfS$ satisfies a closed evolution equation. A direct computation using \eqref{time-Q}--\eqref{time-P}, together with the relations $\bfP^\top \bfQ = 1$ and $\bfP^\top \nabla_x^+ \bfQ = (\nabla_x^- \bfP^\top)\bfQ = 0$, yields
\begin{equation}\label{time-S}
\partial_t \bfS = [\partial_x^2 \bfS, \bfS].
\end{equation}
Indeed,
\begin{equation}
\begin{split}
[\partial_x^2 \bfS,\bfS]
&= \partial_x [\partial_x \bfS,\bfS] \\
&= \partial_x \Big[(\nabla_x^+ \bfQ)\bfP^\top + \bfQ(\nabla_x^- \bfP^\top),\, \bfQ \bfP^\top \Big] \\
&= \partial_x \big((\nabla_x^+ \bfQ)\bfP^\top - \bfQ(\nabla_x^- \bfP^\top)\big) \\
&= \big((\nabla_x^+)^2 \bfQ\big)\bfP^\top - \bfQ\big((\nabla_x^-)^2 \bfP^\top\big) \\
&= \partial_t \bfS,
\end{split}
\end{equation}
where each step follows from straightforward algebra combined with the orthogonality relations \eqref{PdQ} and the equations of motion for $\bfQ$ and $\bfP$.
Let us now define
\begin{equation}
\sigma = 2\bfS - \mathbb{I}.
\end{equation}
Using $\bfS^2=\bfS$, one immediately finds
\[
\sigma^2 = 4\bfS^2 - 4\bfS + \mathbb{I} = \mathbb{I}.
\]
In terms of $\sigma$, the evolution equation \eqref{time-S} becomes the matrix Landau--Lifshitz (or classical Heisenberg ferromagnet) equation
\begin{equation}\label{landau-eq}
\partial_t \sigma = \frac{1}{2}[\partial_x^2 \sigma, \sigma].
\end{equation}
This identifies the multispecies MFT dynamics with a well-known integrable system.
%
The integrability of \eqref{landau-eq} is made explicit by its Lax representation. The associated Lax pair $U(x,t,\eta)$ and $V(x,t,\eta)$, depending on the spectral parameter $\eta$, is given by \cite{takhtajan1977integration,fordy1983nonlinear,faddeev1987hamiltonian}
\[
U = i\eta \sigma,
\qquad
V = 2\eta^2 \sigma - i\eta \sigma \partial_x \sigma.
\]
The zero-curvature condition
\[
\partial_t U - \partial_x V + [U,V]=0
\]
then reproduces the equation of motion \eqref{landau-eq}: the $\eta^2$ terms cancel identically, while the terms linear in $\eta$ yield the Landau--Lifshitz dynamics.

\medskip
\noindent
The main difficulty with eq.~\eqref{landau-eq} lies in the mixed boundary conditions \eqref{BCs0}--\eqref{BCsT}, which are cumbersome to implement within the Landau--Lifshitz formulation. A more convenient representation is obtained by reformulating the problem in AKNS form. This is achieved via the Zakharov--Takhtajan gauge transformation \cite{zakharov1979equivalence}, which maps the state-dependent Lax operator $U=i\eta\sigma(x,t)$ into the standard AKNS structure: a constant diagonal part carrying the spectral parameter, supplemented by an off-diagonal potential encoding the dynamical fields.
To implement this construction, we introduce a local gauge matrix $G(x,t)$ that diagonalizes $\sigma$ into a constant matrix $\Sigma$:
\begin{equation}
G^{-1} \sigma G = \Sigma :=
\begin{pmatrix}
1 & 0 \\
0 & -\mathbb{I}_{N}
\end{pmatrix}.
\end{equation}
The matrix $G$ is built from a complete set of eigenvectors of $\sigma$. Since
$
\sigma = 2\mathbf{Q}\mathbf{P}^\top - \mathbb{I}_{N+1},
$
its spectral structure is particularly simple:
\begin{itemize}
\item the eigenvector associated with eigenvalue $+1$ is $\mathbf{Q}$, using $\mathbf{P}^\top \mathbf{Q}=1$;
\item the $N$ eigenvectors associated with eigenvalue $-1$ are vectors $\mathbf{E}_k$ orthogonal to $\mathbf{P}$, i.e.
\[
\mathbf{P}^\top \mathbf{E}_k = 0.
\]
\end{itemize}
We assemble these vectors into the $(N+1)\times(N+1)$ matrix
\[
G = \begin{pmatrix} \mathbf{Q} & \mathbf{E} \end{pmatrix},
\]
where $\mathbf{E}$ denotes the $(N+1)\times N$ block of negative-eigenvalue eigenvectors. The inverse matrix can be written as
\[
G^{-1} =
\begin{pmatrix}
\mathbf{P}^\top \\
\tilde{\mathbf{E}}^\top
\end{pmatrix},
\]
where $\tilde{\mathbf{E}}$ is the dual basis defined by
\[
\tilde{\mathbf{E}}^\top \mathbf{Q} = 0,
\qquad
\tilde{\mathbf{E}}^\top \mathbf{E} = \mathbb{I}_{N}.
\]

\medskip
\noindent
Under the gauge transformation $\Psi \mapsto \tilde{\Psi}:=G^{-1}\Psi$, the Lax pair transforms as a standard gauge connection:
\begin{equation}
\begin{split}
\tilde{U} &= G^{-1}UG - G^{-1}\partial_x G,\\
\tilde{V} &= G^{-1}VG - G^{-1}\partial_t G.
\end{split}
\end{equation}
We first analyze the spatial operator. Using $G^{-1}\sigma G=\Sigma$, one obtains
\begin{equation}
\tilde{U} = i\eta \Sigma - A,
\qquad
A := G^{-1}\partial_x G,
\end{equation}
so that $A$ plays the role of the AKNS potential.
A block decomposition gives
\begin{equation}
A =
\begin{pmatrix}
\mathbf{P}^\top \partial_x \mathbf{Q} &
\mathbf{P}^\top \partial_x \mathbf{E} \\
\tilde{\mathbf{E}}^\top \partial_x \mathbf{Q} &
\tilde{\mathbf{E}}^\top \partial_x \mathbf{E}
\end{pmatrix}.
\end{equation}
At this stage, two simplifications are available. First, the freedom in the choice of the basis $\mathbf{E}$ can be used to set the $N\times N$ block $\tilde{\mathbf{E}}^\top \partial_x \mathbf{E}$ to zero, fixing a convenient gauge within the $-1$ eigenspace. Second, relation \eqref{PdQ} implies $\mathbf{P}^\top \partial_x \mathbf{Q}=k_J$, which can itself be eliminated by a residual gauge transformation $\mathbf{Q}\to e^{-\Lambda}\mathbf{Q}$, $\mathbf{P}\to e^{\Lambda}\mathbf{P}$, setting $k_J=0$.
After these gauge choices, the matrix $A$ becomes purely off-diagonal:
\begin{equation}
A =
\begin{pmatrix}
0 & \mathbf{q}^\top \\
\mathbf{r} & 0
\end{pmatrix},
\end{equation}
where the AKNS fields are defined by
\begin{equation}\label{canonical}
\mathbf{q}^\top = -\mathbf{E}^\top \partial_x \mathbf{P},
\qquad
\mathbf{r} = \tilde{\mathbf{E}}^\top \partial_x \mathbf{Q}.
\end{equation}
This is the standard AKNS structure, in which the dynamics is fully encoded in off-diagonal potentials.
It remains to understand the effect of this gauge fixing on the boundary conditions. Because $\mathbf{Q}(x,0)$ is piecewise constant with a discontinuity at the origin, the associated gauge field $k_J(t=0,x)$ contains a delta-function contribution. Eliminating this singularity requires a gauge transformation with a discontinuous parameter of the form
\[
\Lambda(t=0,x)=\overline{\Lambda}\,\theta(x).
\]
A similar argument at time $t=T$, together with the assumption $\partial_t\Lambda(x\to\pm\infty)=0$ (which can be justified a posteriori by consistency of the saddle-point equations), leads to the same functional form
\[
\Lambda(t=T,x)=\overline{\Lambda}\,\theta(x).
\]
As a consequence, the boundary conditions are modified into
\begin{subequations}
\begin{align}\label{BCs0L}
\mathbf{Q}(x,0) &= \mathbf{Q}_R e^{-\overline{\Lambda}}\, \theta(x)+\mathbf{Q}_L\,\theta(-x),\\
\label{BCsTL}
\mathbf{P}(x,T) &= \mathbf{P}_R e^{\overline{\Lambda}}\, \theta(x)+\mathbf{P}_L\,\theta(-x),
\end{align}
\end{subequations}
where the remaining constant parameter $\overline{\Lambda}$ encodes the residual global gauge freedom and will be fixed by the saddle-point conditions.

\medskip
\noindent
Substituting $A$ back into the transformed spatial Lax matrix, we recover the characteristic AKNS representation of the multicomponent NLS hierarchy:
\begin{equation}
\tilde{U} =
\begin{pmatrix}
i\eta & -\mathbf{q}^\top \\
-\mathbf{r} & -i\eta \mathbb{I}_{N}
\end{pmatrix}.
\end{equation}
The corresponding transformed temporal matrix $\tilde{V}$ can be computed following the standard AKNS procedure \cite{zakharov1979equivalence} (see Appendix~\ref{gauge-appendix}). One obtains
\begin{equation}
\tilde V(x,t;\eta) =
\begin{pmatrix}
2\eta^2 + \mathbf{q}^\top\mathbf{r}
&
2i\eta \mathbf{q}^\top + \partial_x \mathbf{q}^\top
\\
2i\eta \mathbf{r} - \partial_x \mathbf{r}
&
-2\eta^2 \mathbb{I}_{N} - \mathbf{r}\mathbf{q}^\top
\end{pmatrix}.
\end{equation}
Imposing the zero-curvature condition
\[
\partial_t \tilde{U} - \partial_x \tilde{V} + [\tilde{U}, \tilde{V}] = 0
\]
yields the imaginary-time multicomponent nonlinear Schrödinger (NLS) equations for the AKNS fields:
\begin{subequations}
\begin{align}
\partial_t \mathbf{q} &= -\partial_x^2 \mathbf{q} + 2(\mathbf{q}^\top \mathbf{r})\,\mathbf{q},\\
\partial_t \mathbf{r} &= \partial_x^2 \mathbf{r} - 2(\mathbf{q}^\top \mathbf{r})\,\mathbf{r}.
\end{align}
\end{subequations}
These equations describe a coupled focusing NLS dynamics in imaginary time, with the nonlinear interaction governed entirely by the scalar density $\mathbf{q}^\top\mathbf{r}$.
The boundary conditions inherited from $\mathbf{Q}$ and $\mathbf{P}$, eqs.~\eqref{BCsT}--\eqref{BCs0}, together with the definitions \eqref{canonical}, imply following boundary conditions for the  AKNS fields
\begin{equation}\label{discontinuity2}
\mathbf{r}(x,0)=\overline{\mathbf{r}}\,\delta(x),
\qquad
\mathbf{q}(x,T)=\overline{\mathbf{q}}\,\delta(x).
\end{equation}
%
At this stage, it is important to emphasize that the individual vectors $\overline{\mathbf{r}}$ and $\overline{\mathbf{q}}$ are not gauge-invariant quantities and therefore carry no physical significance on their own. The relevant invariant information is instead contained in the scalar combination
\[
\omega = \overline{\mathbf{q}}^\top \overline{\mathbf{r}},
\]
which, as we will show below, coincides with the single scalar variable $\omega$ already introduced in eq.~\eqref{omega-first}.

\subsection{Linearization of the dynamics}\label{linearization-sect}

\medskip
\noindent
The starting point of the inverse scattering construction is the auxiliary linear problem associated with the Lax pair $(\tilde U,\tilde V)$. It consists of the system
\begin{subequations}
\begin{align}
\label{linear-x}
\partial_x \Psi(x,t;\eta) &= \tilde U(x,t;\eta)\,\Psi(x,t;\eta),\\
\label{linear-t}
\partial_t \Psi(x,t;\eta) &= \tilde V(x,t;\eta)\,\Psi(x,t;\eta),
\end{align}
\end{subequations}
whose compatibility condition is precisely the zero-curvature equation. In this formulation, the nonlinear dynamics of the fields $\mathbf q$ and $\mathbf r$ is encoded into the linear evolution of the auxiliary wavefunction $\Psi$. 

\medskip
\noindent
We assume that the fields $\mathbf q(x,t)$ and $\mathbf r(x,t)$ decay sufficiently rapidly as $|x|\to\infty$. Under this assumption, the Lax matrices asymptotically approach constant diagonal matrices:
\begin{subequations}
\begin{align}
\tilde U(x,t;\eta)
&\xrightarrow[x\to\pm\infty]{}
\tilde U_\infty :=
\begin{pmatrix}
i\eta & 0\\
0 & -i\eta \mathbb I_N
\end{pmatrix},
\\
\tilde V(x,t;\eta)
&\xrightarrow[x\to\pm\infty]{}
\tilde V_\infty :=
\begin{pmatrix}
2\eta^2 & 0\\
0 & -2\eta^2 \mathbb I_N
\end{pmatrix}.
\end{align}
\end{subequations}
This asymptotic structure is essential: far away from the interaction region, the system behaves as a free linear problem, making it possible to define scattering data in complete analogy with ordinary quantum-mechanical scattering theory.

\medskip
\noindent
We now introduce the propagator $\Phi(x,y,t;\eta)$, defined as the fundamental matrix solution of the spatial problem:
\begin{equation}
\label{eq-prop}
\partial_x \Phi(x,y,t;\eta)
=
\tilde U(x,t;\eta)\,\Phi(x,y,t;\eta),
\qquad
\Phi(x,x,t;\eta)=\mathbb I_{N+1}.
\end{equation}
The propagator transports a solution of the auxiliary problem from the point $y$ to the point $x$.
A first important property is that the derivative of $\Phi$ with respect to its second argument also satisfies the same differential equation in $x$. By uniqueness of solutions, it must therefore be proportional to $\Phi$ itself, so that
\begin{equation}
\partial_y \Phi(x,y,t;\eta)
=
\Phi(x,y,t;\eta)\,\tilde U'(y,t;\eta),
\end{equation}
for some matrix $\tilde U'(y,t;\eta)$.
To determine this matrix, we evaluate the above expression at $x=y$. Using the normalization condition $\Phi(y,y,t;\eta)=\mathbb I$, we obtain
\[
\tilde U'(y,t;\eta)
=
-\tilde U(y,t;\eta),
\]
which finally gives
\begin{equation}
\partial_y \Phi(x,y,t;\eta)
=
-\Phi(x,y,t;\eta)\,\tilde U(y,t;\eta).
\end{equation}
Thus the propagator evolves with opposite generators in its two spatial arguments, exactly as expected for a parallel transport operator.

\medskip
\noindent
We now turn to the time evolution of the propagator. Using the zero-curvature condition, one verifies that the quantity
\[
\partial_t\Phi(x,y,t;\eta)
-
\tilde V(x,t;\eta)\Phi(x,y,t;\eta)
\]
also satisfies the same spatial equation as $\Phi$. By the same uniqueness argument, it must therefore be proportional to $\Phi$:
\begin{equation}
\partial_t\Phi(x,y,t;\eta)
-
\tilde V(x,t;\eta)\Phi(x,y,t;\eta)
=
\Phi(x,y,t;\eta)\,M(y,t;\eta),
\end{equation}
for some matrix $M(y,t;\eta)$ depending only on the second argument.
To determine $M$, we again set $x=y$ and use $\Phi(y,y,t;\eta)=\mathbb I$. This immediately yields
\[
M(y,t;\eta)
=
-\tilde V(y,t;\eta).
\]
Substituting back, we arrive at the evolution equation for the propagator:
\begin{equation}\label{evol-prop}
\partial_t\Phi(x,y,t;\eta)
=
\tilde V(x,t;\eta)\Phi(x,y,t;\eta)
-
\Phi(x,y,t;\eta)\tilde V(y,t;\eta).
\end{equation}
Equation \eqref{evol-prop} expresses the fact that the propagator evolves covariantly under the time flow generated by the Lax pair. This relation plays a central role in deriving the evolution of the scattering data and, ultimately, in linearizing the nonlinear dynamics of the original MFT problem.


\medskip
\noindent
We now define the scattering matrix by comparing the asymptotic behavior of the propagator at $x\to+\infty$ and $y\to-\infty$:
\begin{equation}
\label{scatt-matr-def}
S(t,\eta)
=
\lim_{\substack{x\to+\infty\\y\to-\infty}}
e^{-\tilde U_\infty x}\,
\Phi(x,y,t;\eta)\,
e^{\tilde U_\infty y}.
\end{equation}
This matrix encodes how incoming modes from $x=-\infty$ are transformed into outgoing modes at $x=+\infty$.
Using \eqref{evol-prop} and the fact that $\tilde U_\infty$ is time-independent, one finds
\begin{equation}
\partial_t S(t,\eta)
=
\tilde V_\infty\, S(t,\eta)
-
S(t,\eta)\,\tilde V_\infty.
\end{equation}
For our $(N+1)$-component system, it is natural to decompose $S$ as
\begin{equation}
\label{scatt-evol}
S(t,\eta)
=
\begin{pmatrix}
a(t,\eta) & \mathbf b^\top(t,\eta)\\
\overline{\mathbf b}(t,\eta) & \mathcal T(t,\eta)
\end{pmatrix},
\end{equation}
where $a$ is a scalar transmission coefficient, $\mathcal T$ an $N\times N$ transmission matrix, and $\mathbf b$, $\overline{\mathbf b}$ are reflection coefficients. 
The time evolution then reads
\begin{equation}
\partial_t S =
4\eta^2
\begin{pmatrix}
0 & \mathbf b^\top\\
-\overline{\mathbf b} & 0
\end{pmatrix},
\end{equation}
which implies
\begin{subequations}
\begin{align}\label{evol-a}
a(t,\eta) &= a(0,\eta),\\
\label{evol-T}
\mathcal T(t,\eta) &= \mathcal T(0,\eta),\\
\label{evol-b}
\mathbf b(t,\eta) &= \mathbf b(0,\eta)\,e^{4\eta^2 t},\\
\label{evol-bbar}
\overline{\mathbf b}(t,\eta) &= \overline{\mathbf b}(0,\eta)\,e^{-4\eta^2 t}.
\end{align}
\end{subequations}
Thus, the transmission data are conserved, while the reflection data evolve exponentially.
The scattering matrix at $t=0$ and $t=T$ are computed in appendix \ref{scatt-app} with the following results. At time $t=0$ we get
\begin{equation}
S(t=0,\eta)=
\begin{pmatrix} 
1 + \hat{\mathbf{q}}_R^\top \overline{\mathbf{r}} 
& -\hat{\mathbf{q}}_L^\top - \hat{\mathbf{q}}_R^\top -(\hat{\mathbf{q}}^\top_R\overline{\mathbf{r}})\hat{\mathbf{q}}_L^\top \\ 
-\overline{\mathbf{r}} 
& \mathbb{I}_N + \overline{\mathbf{r}}\,\hat{\mathbf{q}}_L^\top 
\end{pmatrix}.
\end{equation}
where we recall $\overline{\mathbf{r}}$ is the vector coefficient appearing in  eq.\eqref{discontinuity2} and we have introduced the Fourier-type transforms
\begin{equation}
\hat{\mathbf{q}}_L(\eta)=\int_{-\infty}^0 e^{-2i\eta y}\mathbf{q}_0(y)\,dy,\qquad
\hat{\mathbf{q}}_R(\eta)=\int_{0}^{\infty} e^{-2i\eta y}\mathbf{q}_0(y)\,dy,
\end{equation}
with $\mathbf{q}_0(y)=\mathbf{q}(y,t=0)$.
Similarly the scattering matrix at time $t=T$ reads
\begin{equation}
S(t=T,\eta)=
\begin{pmatrix} 
1 + \overline{\mathbf{q}}^\top \hat{\mathbf{r}}_L  
& -\overline{\mathbf{q}}^\top\\ 
-\hat{\mathbf{r}}_L - \hat{\mathbf{r}}_R -(\overline{\mathbf{q}}^\top\hat{\mathbf{r}}_L)\hat{\mathbf{r}}_R  
& \mathbb{I}_N + \hat{\mathbf{r}}_R \overline{\mathbf{q}}^\top 
\end{pmatrix}.
\end{equation}
where we recall $\overline{\mathbf{q}}$ is the coefficient appearing in eq.\eqref{discontinuity2} and we have introduced
\begin{equation}
\hat{\mathbf{r}}_L(\eta)=\int_{-\infty}^0 e^{2i\eta y}\mathbf{r}_T(y)\,dy,\qquad
\hat{\mathbf{r}}_R(\eta)=\int_{0}^{\infty} e^{2i\eta y}\mathbf{r}_T(y)\,dy,
\end{equation}
with $\mathbf{r}_T(y)=\mathbf{r}(y,t=T)$.
Now we use the evolution equation of the scattering matrix eqs.\eqref{evol-a}--\eqref{evol-bbar}.
From eq.\eqref{evol-T} we get
\begin{equation}\label{qL-rR}
\hat{\mathbf{q}}_L(\eta) =\hat v_+(\eta)\,\overline{\mathbf{q}},\qquad \hat{\mathbf{r}}_R(\eta) =\hat v_+(\eta)\,\overline{\mathbf{r}}.
\end{equation}
Eqs.\eqref{evol-b}--\eqref{evol-bbar} read
\begin{equation}\label{off-diagonal}
\begin{split}
\overline{\mathbf{q}}e^{-4\eta^2 T} 
&= 
\hat{\mathbf{q}}_L(\eta)+\hat{\mathbf{q}}_R(\eta)+ (\overline{\mathbf{r}}^\top\hat{\mathbf{q}}_R)\hat{\mathbf{q}}_L(\eta)
\\
\overline{\mathbf{r}}e^{-4\eta^2 T} 
&=
\hat{\mathbf{r}}_L(\eta)+\hat{\mathbf{r}}_R(\eta)+ (\overline{\mathbf{q}}^\top\hat{\mathbf{r}}_L)\hat{\mathbf{r}}_R(\eta)
\end{split}
\end{equation}
Given eq.~\eqref{qL-rR}, we know that $\hat{\mathbf q}_L(\eta)$ is proportional to $\overline{\mathbf q}$ and that $\hat{\mathbf r}_R(\eta)$ is proportional to $\overline{\mathbf r}$. The previous relations then imply that $\hat{\mathbf q}_R(\eta)$ and $\hat{\mathbf r}_L(\eta)$ are proportional to the same vectors, respectively. Moreover, using eq.~\eqref{evol-a}, one finds that the corresponding proportionality coefficients coincide. We may therefore write
\begin{equation}\label{qR-rL}
\hat{\mathbf q}_R(\eta)
=
\hat v_-(\eta)\,\overline{\mathbf q},
\qquad
\hat{\mathbf r}_L(\eta)
=
\hat v_-(\eta)\,\overline{\mathbf r}.
\end{equation}
Collecting the different contributions, we conclude that the boundary fields factorize as
\begin{equation}
\mathbf r_T(x)
=
\overline{\mathbf r}\,v(x),
\qquad
\mathbf q_0(x)
=
\overline{\mathbf q}\,v(-x),
\end{equation}
where the scalar function $v(x)$ is characterized through its half-space Fourier transforms
\begin{equation}
\hat v_+(\eta)
=
\int_0^\infty dx \, e^{2i\eta x}v(x),
\qquad
\hat v_-(\eta)
=
\int_{-\infty}^0 dx \, e^{2i\eta x}v(x).
\end{equation}
The matrix relations \eqref{off-diagonal} then reduce to the single scalar equation
\begin{equation}\label{RH}
e^{-4\eta^2T}
=
\hat v_+(\eta)
+
\hat v_-(\eta)
+
\omega\,\hat v_+(\eta)\hat v_-(\eta),
\end{equation}
where
\[
\omega
=
\overline{\mathbf q}^{\top}\overline{\mathbf r}.
\]
Up to a minor difference in normalization, this is precisely the scalar equation obtained in \cite{mallick2022exact}, and which had previously appeared in a disguised form in \cite{grabsch2022exact}. Similar equations also arise in the inverse-scattering solution of the MFT equations for the KMP model \cite{bettelheim2022inverse}.

Equation \eqref{RH} is a scalar Riemann--Hilbert problem: one seeks a factorization of a given analytic function into components analytic in the upper and lower half-planes. Its solution is known explicitly \cite{mallick2022exact}:
\begin{equation}\label{RH-solution}
\hat v_\pm(\eta)
=
\frac{e^{\Phi_\pm(\eta)}-1}{\omega},
\end{equation}
with
\begin{equation}
\Phi_\pm(\eta)
=
\lim_{\epsilon\to0^+}
\pm\frac{1}{2\pi i}
\int_{-\infty}^{+\infty}
\frac{
\log\!\left(1+\omega e^{-4(q+\eta)^2T}\right)
}{
q\mp i\epsilon
}\,dq.
\end{equation}

\subsection{Initial and final density profile and particle transfer}
\label{sect-profile}

Once the Riemann--Hilbert problem has been solved, the optimal density profiles can be reconstructed explicitly.
To do so, we construct the matrix $G$ associated with the Zakharov--Takhtajan gauge transformation, together with its inverse $G^{-1}$. 
This construction provides direct access to the vectors $\mathbf{Q}$ and $\mathbf{P}$, and therefore to the density $\boldsymbol{\rho}$.

\medskip

\noindent
We parametrize $G$ and its inverse as
\begin{equation}
G = \begin{pmatrix} 
\mathbf{Q} & \mathbf{E} 
\end{pmatrix},
\qquad
G^{-1} = 
\begin{pmatrix} 
\mathbf{P}^\top \\ \tilde{\mathbf{E}}^\top 
\end{pmatrix}.
\end{equation}
By construction, these matrices satisfy the first-order differential equations (see Eq.~\eqref{A-G})
\begin{equation}
\begin{split}
\partial_x G(x,t) &= G(x,t)\,A(x,t),\\[6pt]
\partial_x G^{-1}(x,t) &= -A(x,t)\,G^{-1}(x,t).
\end{split}
\end{equation}
Hence, once appropriate boundary conditions are specified, these equations uniquely determine $G$ and $G^{-1}$.

\medskip
\noindent
$\bullet$ \textbf{Initial time $t=0$.}

\medskip
\noindent
At $t=0$, the vector $\mathbf{Q}(x,0)$ is known (see eq.~\eqref{BCs0L}). 
Our task is therefore to determine $\mathbf{P}(x,0)$ from the equation for $G^{-1}$.
At this time, the matrix $A$ takes the form
\begin{equation}
A(x,0)=\begin{pmatrix} 
0 & \mathbf{q}_0^\top(x) \\ 
\overline{\mathbf{r}}\,\delta(x) & 0 
\end{pmatrix}.
\end{equation}
Substituting into the evolution equation for $G^{-1}$ gives
\begin{equation}
\begin{pmatrix} 
\partial_x \mathbf{P}^\top \\ 
\partial_x \tilde{\mathbf{E}}^\top  
\end{pmatrix}
=
-\begin{pmatrix} 
0 & \mathbf{q}_0^\top(x) \\ 
\overline{\mathbf{r}}\,\delta(x) & 0 
\end{pmatrix}
\begin{pmatrix} 
\mathbf{P}^\top \\ 
\tilde{\mathbf{E}}^\top  
\end{pmatrix}.
\end{equation}
Carrying out the matrix multiplication, we obtain the coupled system
\begin{equation}
\begin{split}
\partial_x \mathbf{P}^\top &= -\mathbf{q}_0^\top(x)\,\tilde{\mathbf{E}}^\top,\\
\partial_x \tilde{\mathbf{E}}^\top &= -\overline{\mathbf{r}}\,\mathbf{P}^\top\,\delta(x).
\end{split}
\end{equation}

\medskip
\noindent
We now integrate these equations separately on the two half-lines $(-\infty,0)$ and $(0,\infty)$, taking into account the discontinuity at $x=0$. 
Denoting $\mathbf{P}_0 = \mathbf{P}(0)$, we obtain
\begin{equation}
\begin{split}
\mathbf{P}_R e^{\overline{\Lambda}} - \mathbf{P}_0 
&= -\tilde{\mathbf{E}}_R \int_0^\infty dx\, \mathbf{q}_0(x),\\
\mathbf{P}_0 - \mathbf{P}_L 
&= -\tilde{\mathbf{E}}_L \int_{-\infty}^0 dx\, \mathbf{q}_0(x),
\end{split}
\end{equation}
together with the jump condition
\begin{equation}
\tilde{\mathbf{E}}_L = \tilde{\mathbf{E}}_R + \mathbf{P}_0\,\overline{\mathbf{r}}^\top,
\end{equation}
which follows from integrating the $\delta$-term across the origin.

\medskip

\noindent
Using the definitions of $\hat v_\pm(0)$, this system becomes
\begin{subequations}
\begin{align}
\mathbf{P}_Re^{\overline{\Lambda}} - \mathbf{P}_0
&= -\tilde{\mathbf{E}}_R \overline{\mathbf{q}}\,\hat v_-(0),\\
\mathbf{P}_0 - \mathbf{P}_L
&= -\left(\tilde{\mathbf{E}}_R \overline{\mathbf{q}} + \omega \mathbf{P}_0\right)\hat v_+(0).
\end{align}
\end{subequations}
This is a linear system for the two unknowns $\mathbf{P}_0$ and $\tilde{\mathbf{E}}_R \overline{\mathbf{q}}$, which can be solved explicitly. 
The solution reads
\begin{subequations}
\begin{align}
\tilde{\mathbf{E}}_R \overline{\mathbf{q}}
&= \mathbf{P}_L - \sqrt{1+\omega}\,\mathbf{P}_R e^{\overline{\Lambda}},\\
\mathbf{P}_0
&= \frac{1}{\omega}(\mathbf{P}_L+\mathbf{P}_R e^{\overline{\Lambda}})(\sqrt{1+\omega}-1).
\end{align}
\end{subequations}
Next, we use the normalization condition $\mathbf{P}\cdot \mathbf{Q}=1$. 
Evaluating it on both sides of the origin gives
\begin{equation}
\mathbf{P}_0 \cdot \mathbf{Q}_R\,e^{-\overline{\Lambda}} =
\mathbf{P}_0 \cdot \mathbf{Q}_L= 1.
\end{equation}
These relations allow $\overline{\Lambda}$
 as
\begin{equation}
e^{2\overline{\Lambda}}=
\frac{\bfP_L \cdot \bfQ_R}{\bfP_R \cdot \bfQ_L}=
\frac{\sum_{\alpha=0}^{N}\rho_{R,\alpha}e^{-\lambda_\alpha}}
{\sum_{\alpha=0}^{N}\rho_{L,\alpha}e^{\lambda_\alpha}},
\end{equation}
as well as the variable $\omega$
\begin{equation}
\omega = 
(\bfP_R \cdot \bfQ_L)(\bfP_L \cdot \bfQ_R)-1
=
\left(\sum_{\alpha=0}^{N}\rho_{L,\alpha}e^{\lambda_\alpha}\right)
\left(\sum_{\alpha=0}^{N}\rho_{R,\alpha}e^{-\lambda_\alpha}\right) - 1.
\end{equation}
Finally, integrating the differential equation for $\mathbf{P}$ gives
\begin{equation}
\mathbf{P}(x,0)
=
\begin{cases}
\mathbf{P}_R \, e^{\overline{\Lambda}} + (\mathbf{P}_L - \sqrt{1+\omega}\mathbf{P}_R\, e^{\overline{\Lambda}})
\displaystyle\int_x^\infty dy\, v(-y), & x>0,\\[10pt]
\mathbf{P}_L + (\mathbf{P}_R\, e^{\overline{\Lambda}} - \sqrt{1+\omega}\mathbf{P}_L)
\displaystyle\int_{-\infty}^x dy\, v(-y), & x<0.
\end{cases}
\end{equation}
This immediately yields the initial density profile
\begin{equation}
\rho_\alpha(x,0)=
\begin{cases}
\rho_{R,\alpha}\left[1+
\left(e^{-\lambda_\alpha}e^{-\overline{\Lambda}}-\sqrt{1+\omega}\right)
\displaystyle\int_x^\infty dy\, v(-y)\right], & x>0,\\[10pt]
\rho_{L,\alpha}\left[1+
\left(e^{\lambda_\alpha}e^{\overline{\Lambda}}-\sqrt{1+\omega}\right)
\displaystyle\int_{-\infty}^x dy\, v(-y)\right], & x<0.
\end{cases}
\end{equation}

\medskip
\noindent
$\bullet$ \textbf{Final time $t=T$.}

\medskip
\noindent
The derivation at final time follows the same steps. 
Now $\mathbf{P}(x,T)$ is known, and we determine $\mathbf{Q}(x,T)$ from the equation for $G$.
The matrix $A$ is
\begin{equation}
A(x,T)=
\begin{pmatrix} 
0 & \overline{\mathbf{q}}^\top\,\delta(x) \\ 
\mathbf{r}_T(x) & 0 
\end{pmatrix}.
\end{equation}
Repeating the previous analysis, we obtain
\begin{subequations}
\begin{align}
\mathbf{E}_R \overline{\mathbf{r}}
&=
-\mathbf{Q}_L + \sqrt{1+\omega}\,\mathbf{Q}_R \,e^{-\overline{\Lambda}},\\
\mathbf{Q}_0
&=
\frac{1}{\omega}(\mathbf{Q}_L+\mathbf{Q}_R\,\,e^{-\overline{\Lambda}})(\sqrt{1+\omega}-1).
\end{align}
\end{subequations}
Imposing the conditions $1=\bfP_R\cdot \bfQ_0=\bfP_L\cdot \bfQ_0$, returns not only as expected the same expression of $\omega$ found above, 
but also the same expression for $\overline{\Lambda}$, which confirms the assumption that the asymptotic in space value of $\Lambda$ can be taken independent of time.
Thus,
\begin{equation}
\mathbf{Q}(x,T)
=
\begin{cases}
\mathbf{Q}_R \, e^{-\overline{\Lambda}}+ (\mathbf{Q}_L - \sqrt{1+\omega}\mathbf{Q}_R\, e^{-\overline{\Lambda}})
\displaystyle\int_x^\infty dy\, v(y), & x>0,\\[10pt]
\mathbf{Q}_L + (\mathbf{Q}_R\, e^{-\overline{\Lambda}} - \sqrt{1+\omega}\mathbf{Q}_L)
\displaystyle\int_{-\infty}^x dy\, v(y), & x<0,
\end{cases}
\end{equation}
and the final density profile
\begin{equation}
\rho_\alpha(x,T)=
\begin{cases}
\rho_{R,\alpha}+
\left[\rho_{L,\alpha} e^{\lambda_\alpha}e^{\overline{\Lambda}}
-\sqrt{1+\omega}\,\rho_{R,\alpha}\right]
\displaystyle\int_x^\infty dy\, v(y), & x>0,\\[10pt]
\rho_{L,\alpha}+
\left[\rho_{R,\alpha} e^{-\lambda_\alpha}e^{-\overline{\Lambda}}
-\sqrt{1+\omega}\,\rho_{L,\alpha}\right]
\displaystyle\int_{-\infty}^x dy\, v(y), & x<0.
\end{cases}
\end{equation}

\subsubsection{Total transferred particles}
We can now compute the total transferred particles
\begin{equation}
\begin{split}
C_\alpha(T)&= \int_0^\infty dx \,(\rho_\alpha(x,T)-\rho_\alpha(x,0)).
\end{split}
\end{equation}
Using symmetry properties of $v$, one finds
\[
\int_0^\infty dy \,y v(y)=\int_0^\infty dy \,y v(-y)=\frac{1}{2i}\hat{v}'_+(0),
\]
which leads to
\begin{equation}
C_\alpha(T)
=
\left(\rho_{L,\alpha}
 e^{\lambda_\alpha}e^{\Delta\Lambda_0}-\rho_{R,\alpha}e^{-\lambda_\alpha}e^{-\Delta\Lambda_0}
\right)\frac{1}{2i}\hat{v}'_+(0).
\end{equation}
The prefactor can be rewritten as
\[
\rho_{L,\alpha}
 e^{\lambda_\alpha}e^{\Delta\Lambda_0}-\rho_{R,\alpha}
 e^{-\lambda_\alpha}e^{-\Delta\Lambda_0}
=
\frac{\partial \omega}{\partial\lambda_\alpha}\frac{1}{\sqrt{1+\omega}},
\]
so that
\begin{equation}
C_\alpha(T)= 
\frac{\partial \omega}{\partial\lambda_\alpha}
\frac{\hat v'_+(0)}{2i\sqrt{1+\omega}}.
\end{equation}
Using the Riemann--Hilbert solution, one obtains
\begin{equation}
\frac{\hat{v}'_+(0)}{2i\sqrt{1+\omega}}
= 
\frac{\sqrt{T}}{\pi} \int_{-\infty}^{+\infty}
\frac{ e^{- q^2 }}{1+\omega e^{- q^2 }}dq,
\end{equation}
and therefore
\begin{equation}
C_\alpha(T)=\sqrt{T}\frac{\partial \mu(\bflambda)}{\partial \lambda_\alpha}
\quad \text{with}
\quad
\mu(\boldsymbol{\lambda}) = \frac{1}{\pi} \int_{-\infty}^{+\infty} dk\log\left[1+\omega e^{-k^2}  \right].
\end{equation}
This result is fully consistent with the general framework developed in Section~\ref{DG-section}: the cumulant generating function depends on the initial densities and counting fields only through the single scalar combination $\omega$. Moreover, it naturally recovers the known expression for the single-species SSEP \cite{derrida2009current,mallick2022exact}, confirming the consistency of the multicomponent construction.

\section{Conclusion}\label{concl-sect}

We have studied the macroscopic fluctuation theory of the multispecies symmetric simple exclusion process on the infinite line, working throughout with the redundant set of densities $\bfrho=\{\rho_0,\dots,\rho_N\}$ in order to keep the relabelling symmetry of the model manifest. We showed that the saddle-point equations of MFT are of Landau--Lifshitz type and, after a Zakharov--Takhtajan gauge transformation, can be cast in AKNS form; this identifies the underlying integrable structure and allows the inverse scattering method to be used to solve them explicitly. As a consequence, we obtained the cumulant generating function of the multispecies current, recovering its dependence on a single scaling variable $\omega$, together with the density profiles conditioned on a prescribed current fluctuation. These results extend to the multispecies setting the integrability-based approach previously developed for the single-species SSEP \cite{derrida2009current,mallick2022exact}.

Several extensions of this work suggest themselves. A natural one is the weakly asymmetric exclusion process (WASEP),
for which it is plausible that the multispecies WASEP saddle-point equations retain an analogous Lax structure, allowing the inverse scattering approach developed here to be adapted to the weakly asymmetric, multispecies setting. Establishing this would in particular clarify how species-dependent drifts affect the current statistics, and whether the resulting large deviation function still depends on the boundary data through a single scaling variable as in~\eqref{omega-dep}.

A more speculative, but potentially far-reaching, direction concerns systems with boundaries, in the spirit of recent work on boundary-driven models \cite{suzuki2026non}. There, MFT must be supplemented with boundary conditions coupling the system to reservoirs, and it is at present unclear whether the resulting saddle-point problem retains an integrable structure analogous to the one uncovered here, or whether the Lax pair and inverse scattering method need to be substantially reformulated to accommodate the boundary terms. Settling this question --- in particular, identifying the correct boundary conditions on the scattering data and determining whether they preserve solvability by inverse scattering --- would be an important step towards a unified, integrability-based treatment of multispecies diffusive fluctuations beyond the infinite-line geometry considered here.

\section*{Acknowledgments}

We thank Kirone Mallick for suggesting the references \cite{bodineau2011phase,bodineau2008long}.

\medskip
\noindent
\textbf{Use of AI tools.} During the preparation of this manuscript, the author used the large language model Claude (Claude Sonnet 4.6, Anthropic) as an assistive tool for three purposes: (i) editing and polishing the text, including improving clarity, grammar, and consistency of notation across sections; (ii) helping to identify and locate relevant literature on macroscopic fluctuation theory, multispecies exclusion processes, and integrable systems; and (iii) cross-checking some of the intermediate calculations. All such LLM-assisted checks were independently verified by the author through his own derivations, and every calculation and result presented in the paper rests on this independent verification rather than on the LLM output alone. All scientific content, derivations, and conclusions are the author's own, and all AI-assisted text and suggested references were likewise carefully checked by the author before inclusion. The author take full responsibility for the content of the manuscript.

\appendix

\addcontentsline{toc}{section}{Appendix}

\section{A simple example}\label{simple-ex}


We present a very simple pedagogical example which illustrates the general
structure of the cumulant generating function of the integrated currents.
Consider a graph consisting of only two vertices, denoted by $L$ and $R$.
Initially each vertex $i\in\{L,R\}$ is occupied by a particle of species $\alpha$
with probability $\rho_{i,\alpha}$. 
We assume that the initial measure is factorized, so that the joint
probability of the configuration $(\alpha,\beta)$ (species $\alpha$ at site $a$ and
species $\beta$ at site $b$) is
\begin{equation}
P_{0}(\alpha,\beta)= \rho_{L,\alpha}\rho_{R,\beta}.
\end{equation}
%
%
Particles of different species exchange their positions with rate $1$.
Therefore the only possible transition is
\[
(\alpha,\beta) \longleftrightarrow (\beta,\alpha).
\]
Given that the system starts in configuration $(\alpha,\beta)$,
it is straightforward to compute the probability of being in
configuration $(\alpha,\beta)$ or $(\beta,\alpha)$ at time $t$.
If $\alpha\neq \beta$ we obtain the two--state Markov evolution
\begin{align}
P[(\alpha,\beta),t \mid (\alpha,\beta),0] &=  \frac{1+e^{-2t}}{2}, \\
P[(\beta,\alpha),t \mid (\alpha,\beta),0] &=  \frac{1-e^{-2t}}{2}.
\end{align}
If the two particles are identical ($\alpha=\beta$), no exchange is observable and
the configuration remains unchanged:
\begin{equation}
P[(\alpha,\alpha),t \mid (\alpha,\alpha),0]=1.
\end{equation}
%
%
Multiplying by the probability of the initial configuration gives the joint
probabilities
\begin{align}
P[(\alpha,\beta),t;(\alpha,\beta),0] &=  \frac{1+e^{-2t}}{2}\rho_{L,\alpha}\rho_{R,\beta}, \\
P[(\beta,\alpha),t;(\alpha,\beta),0] &=  \frac{1-e^{-2t}}{2}\rho_{L,\alpha}\rho_{R,\beta},
\end{align}
for $\alpha\neq \beta$, while for $\alpha=\beta$ we have
\begin{equation}
P[(\alpha,\alpha),t;(\alpha,\alpha),0]=\rho_{L,\alpha}\rho_{R,\alpha}.
\end{equation}
%
%
The only evolutions producing a non–zero charge transfer on site $a$
are those in which the two particles exchange their positions:
\[
(\alpha,\beta)\longleftrightarrow (\beta,\alpha),
\qquad \alpha\neq \beta.
\]
In such an event the charge transfers to the vertex $R$ are
\[
C_\alpha = +1,
\qquad
C_\beta = -1,
\]
since a particle of species $\alpha$ enters site $R$ while a particle
of species $\beta$ leaves it.
%
%
The moment generating function therefore reads
\begin{equation}
f(\underline{\lambda};\underline{\bfrho};t)
=
\sum_{\alpha\neq \beta}
\frac{1+e^{-2t}}{2}\rho_{L,\alpha}\rho_{R,\beta}
+
\sum_{\alpha\neq \beta}
\frac{1-e^{-2t}}{2}
\rho_{L,\alpha}e^{\lambda_\alpha}
\rho_{R,\beta}e^{-\lambda_\beta}
+
\sum_\alpha \rho_{L,\alpha}\rho_{R,\alpha}.
\end{equation}
%
%
Using the normalization conditions
\[
\sum_{\alpha=0}^N \rho_{L,\alpha}=
\sum_{\alpha=0}^N \rho_{R,\alpha}=1,
\]
a straightforward algebraic simplification gives
\begin{equation}
f(\underline{\lambda};\underline{\bfrho};t)
=
1+
\left[
\left(
\sum_{\alpha=0}^N \rho_{L,\alpha}e^{\lambda_\alpha}
\right)
\left(
\sum_{\alpha=0}^N \rho_{R,\alpha}e^{-\lambda_\alpha}
\right)-1\right]
\left(\frac{1-e^{-2t}}{2}\right),
\end{equation}
%
%
The generating function depends
on the fugacities and the densities only through the combination
\[
\omega=
\left(\sum_{\alpha} \rho_{L,\alpha}e^{\lambda_\alpha}\right)
\left(\sum_{\alpha} \rho_{R,\alpha}e^{-\lambda_\alpha}\right)-1.
\]
in agreement with  the general structure
obtained in Section \ref{DG-section}.

\section{Zakharov--Takhtajan gauge transformation}\label{gauge-appendix}

\medskip
\noindent
We start from the Lax pair associated with the matrix Landau--Lifshitz equation,
\begin{equation}
U = i\eta\, \sigma,
\qquad
V = 2\eta^2 \sigma - i\eta\, \sigma\, \partial_x \sigma,
\end{equation}
which satisfies the zero-curvature condition
\begin{equation}
\partial_t U - \partial_x V + [U,V] = 0.
\end{equation}
Here $\eta$ is the spectral parameter and the field $\sigma$ is given by
\begin{equation}
\sigma = 2\,\mathbf{Q}\mathbf{P}^\top - \mathbb{I}_{N+1}.
\end{equation}

\medskip
\noindent
In order to bring this Lax pair into the standard AKNS form, we perform a gauge transformation based on the eigenstructure of $\sigma$. We introduce the matrix
\begin{equation}
G = \begin{pmatrix} \mathbf{Q} & \mathbf{E} \end{pmatrix},
\qquad
G^{-1} = \begin{pmatrix} \mathbf{P}^\top \\ \tilde{\mathbf{E}}^\top \end{pmatrix},
\end{equation}
where the vectors satisfy the orthogonality and normalization conditions
\begin{equation}
\mathbf{P}^\top \mathbf{Q} = 1,
\qquad
\mathbf{P}^\top \mathbf{E} = 0,
\qquad
\tilde{\mathbf{E}}^\top \mathbf{Q} = 0,
\qquad
\tilde{\mathbf{E}}^\top \mathbf{E} = \mathbb{I}_N.
\end{equation}

\medskip
\noindent
Under the gauge transformation $\Psi \to \tilde{\Psi} := G^{-1}\Psi$, the Lax pair transforms as
\begin{equation}
\tilde{U} = G^{-1} U G - G^{-1} \partial_x G,
\qquad
\tilde{V} = G^{-1} V G - G^{-1} \partial_t G.
\end{equation}

\medskip
\noindent
Let us first analyze the spatial part. Using the fact that $G^{-1}\sigma G = \Sigma$, where
\begin{equation}
\Sigma =
\begin{pmatrix}
1 & 0\\
0 & -\mathbb{I}_N
\end{pmatrix},
\end{equation}
we obtain
\begin{equation}\label{A-G}
\tilde{U} = i\eta \Sigma - A,
\qquad
A := G^{-1} \partial_x G.
\end{equation}

\medskip
\noindent
The matrix $A$ can be computed explicitly:
\begin{equation}
A =
\begin{pmatrix}
\mathbf{P}^\top\\
\tilde{\mathbf{E}}^\top
\end{pmatrix}
\begin{pmatrix}
\partial_x \mathbf{Q} & \partial_x \mathbf{E}
\end{pmatrix}
=
\begin{pmatrix}
\mathbf{P}^\top \partial_x \mathbf{Q} &
\mathbf{P}^\top \partial_x \mathbf{E}\\
\tilde{\mathbf{E}}^\top \partial_x \mathbf{Q} &
\tilde{\mathbf{E}}^\top \partial_x \mathbf{E}
\end{pmatrix}.
\end{equation}

\medskip
\noindent
At this stage we exploit the gauge freedom. Choosing the gauge
\begin{equation}
\mathbf{P}^\top \partial_x \mathbf{Q} = 0,
\qquad
\tilde{\mathbf{E}}^\top \partial_x \mathbf{E} = 0,
\end{equation}
the matrix $A$ becomes purely off-diagonal. Introducing the fields
\begin{equation}
\mathbf{q}^\top = \mathbf{P}^\top \partial_x \mathbf{E},
\qquad
\mathbf{r} = \tilde{\mathbf{E}}^\top \partial_x \mathbf{Q},
\end{equation}
we obtain
\begin{equation}
A =
\begin{pmatrix}
0 & \mathbf{q}^\top\\
\mathbf{r} & 0
\end{pmatrix}.
\end{equation}

\medskip
\noindent
We now turn to the temporal part $\tilde{V}$. Since the original matrix $V$ is quadratic in $\eta$, it is natural to expand
\begin{equation}
\tilde{V} = \tilde{V}_0 + \eta\, \tilde{V}_1 + \eta^2\, \tilde{V}_2.
\end{equation}

\medskip
\noindent
The highest-order term is immediately obtained:
\begin{equation}
\tilde{V}_2 = G^{-1}(2\sigma)G = 2\Sigma.
\end{equation}

\medskip
\noindent
The linear term involves the derivative of $\sigma$. Using $\sigma = 2\mathbf{Q}\mathbf{P}^\top - \mathbb{I}$, one finds after straightforward algebra
\begin{equation}
\tilde{V}_1 = 2i A.
\end{equation}

\medskip
\noindent
Altogether, the transformed Lax pair takes the form
\begin{equation}
\begin{split}
\tilde{U} &= i\eta \Sigma - A,\\
\tilde{V} &= 2\eta^2 \Sigma + 2i\eta A + \tilde{V}_0.
\end{split}
\end{equation}

\medskip
\noindent
To determine $\tilde{V}_0$, we impose the zero-curvature condition
\begin{equation}
\partial_t \tilde{U} - \partial_x \tilde{V} + [\tilde{U},\tilde{V}] = 0,
\end{equation}
and expand it order by order in $\eta$. This yields
\begin{equation}
\begin{split}
\eta^2 &: \quad [-A,2\Sigma] + [i\Sigma,2iA] = 0,\\
\eta^1 &: \quad -2i\,\partial_x A + [i\Sigma,\tilde{V}_0] = 0,\\
\eta^0 &: \quad -\partial_t A - \partial_x \tilde{V}_0 - [A,\tilde{V}_0] = 0.
\end{split}
\end{equation}

\medskip
\noindent
The $\eta^2$ equation is identically satisfied. To solve the remaining equations, we parametrize
\begin{equation}
\tilde{V}_0 =
\begin{pmatrix}
b & \tilde{\mathbf{m}}^\top\\
\mathbf{m} & \mathbf{B}
\end{pmatrix}.
\end{equation}

\medskip
\noindent
From the $\eta^1$ order, one obtains
\begin{equation}
\mathbf{m} = -\partial_x \mathbf{r},
\qquad
\tilde{\mathbf{m}} = \partial_x \mathbf{q}.
\end{equation}

\medskip
\noindent
From the diagonal part of the $\eta^0$ equation, we find
\begin{equation}
\partial_x b = \partial_x(\mathbf{q}^\top \mathbf{r}),
\qquad
\partial_x \mathbf{B} = -\partial_x(\mathbf{r}\mathbf{q}^\top),
\end{equation}
which, using appropriate boundary conditions at infinity, yields
\begin{equation}
b = \mathbf{q}^\top \mathbf{r},
\qquad
\mathbf{B} = -\mathbf{r}\mathbf{q}^\top.
\end{equation}

\medskip
\noindent
Finally, the off-diagonal part of the $\eta^0$ equation gives the evolution equations for the fields:
\begin{equation}
\partial_t \mathbf{q}
=
-\partial_x^2 \mathbf{q}
+ 2(\mathbf{q}^\top \mathbf{r})\,\mathbf{q},
\end{equation}
\begin{equation}
\partial_t \mathbf{r}
=
\partial_x^2 \mathbf{r}
- 2(\mathbf{q}^\top \mathbf{r})\,\mathbf{r}.
\end{equation}

\medskip
\noindent
These are precisely the (imaginary-time) multicomponent nonlinear Schrödinger (NLS) equations obtained from the AKNS hierarchy.

\section{Scattering matrices}\label{scatt-app}

\medskip
\noindent
To compute the scattering matrix  $S$, we introduce the Jost solutions $\Phi_L$ and $\Phi_R$, defined by their asymptotics:
\begin{equation}
\label{asympt1}
\lim_{x\to-\infty} e^{-\tilde U_\infty x}\Phi_L(x)=\mathbb I,
\qquad
\lim_{x\to+\infty} e^{-\tilde U_\infty x}\Phi_R(x)=\mathbb I.
\end{equation}
They represent solutions normalized at the left and right infinities, respectively.
The propagator factorizes as
\begin{equation}
\Phi(x,y)
=
\Phi_R(x)\Phi_R^{-1}(z)\Phi_L(z)\Phi_L^{-1}(y),
\end{equation}
which implies the standard relation
\begin{equation}
S = \Phi_R^{-1}(z)\Phi_L(z),
\end{equation}
independent of the intermediate point $z$.

\medskip
\noindent
Let us now compute explicitly the scattering data at the initial and final times.  
We begin with the case $t=0$, where the structure of the potential is particularly simple.

\medskip
\noindent
At $t=0$, the spatial Lax matrix takes the form
\begin{equation}
\tilde{U}(x,0;\eta)= 
\begin{pmatrix} 
i\eta & -\mathbf{q}^\top(x,t=0) \\ 
-\overline{\mathbf{r}}\,\delta(x) & -i\eta \mathbb{I}_{N} 
\end{pmatrix}.
\end{equation}
Away from the origin, the delta contribution vanishes, so the system is effectively triangular. This allows us to construct the Jost solutions explicitly on each half-line.

\medskip
\noindent
On the left half-line ($x<0$), the solution $\Phi_L(x)$ is defined by its asymptotic normalization at $x\to -\infty$. Solving the linear problem in this region yields
\begin{equation}
\Phi_L(x) = e^{i\Sigma x}
\begin{pmatrix} 
1 & -\displaystyle\int_{-\infty}^x e^{-2i\eta y} \mathbf{q}^\top(y,t=0)\,dy \\ 
0 & \mathbb{I}_N 
\end{pmatrix}.
\end{equation}
Similarly, on the right half-line ($x>0$), the Jost solution $\Phi_R(x)$, normalized at $x\to +\infty$, is given by
\begin{equation}
\Phi_R(x) = e^{i\Sigma x}
\begin{pmatrix} 
1 & \displaystyle\int_x^{\infty} e^{-2i\eta y} \mathbf{q}^\top(y,t=0)\,dy \\ 
0 & \mathbb{I}_N 
\end{pmatrix}.
\end{equation}

\medskip
\noindent
The scattering matrix is obtained by matching these two solutions at the origin. In principle, one would write
\begin{equation}
S(t=0,\eta)=\Phi_R^{-1}(0^+)\Phi_L(0^+).
\end{equation}
However, because of the delta function in $\tilde U$, the solution $\Phi_L$ is discontinuous at $x=0$. The discontinuity is encoded in a jump condition
\begin{equation}
\Phi_L(0^+)=\mathbf{J}\,\Phi_L(0^-),
\end{equation}
where the jump matrix is
\begin{equation}
\mathbf{J}=
\begin{pmatrix} 
1 & 0 \\ 
-\overline{\mathbf{r}} & \mathbb{I}_N 
\end{pmatrix}.
\end{equation}
Therefore the scattering matrix becomes
\begin{equation}
S(t=0,\eta)=\Phi_R^{-1}(0^+)\,\mathbf{J}\,\Phi_L(0^-).
\end{equation}

\medskip
\noindent
To make this expression explicit, it is convenient to introduce the Fourier-type transforms
\begin{equation}
\hat{\mathbf{q}}_L(\eta)=\int_{-\infty}^0 e^{-2i\eta y}\mathbf{q}(y,t=0)\,dy,\qquad
\hat{\mathbf{q}}_R(\eta)=\int_{0}^{\infty} e^{-2i\eta y}\mathbf{q}(y,t=0)\,dy,
\end{equation}
and
\begin{equation}
\hat{\mathbf{q}}(\eta)=\hat{\mathbf{q}}_L(\eta)+\hat{\mathbf{q}}_R(\eta).
\end{equation}
Evaluating the Jost solutions at the origin gives
\begin{equation}
\Phi_L(0^-)=
\begin{pmatrix} 
1 & -\hat{\mathbf{q}}_L^\top(\eta) \\ 
0 & \mathbb{I}_N 
\end{pmatrix},
\qquad
\Phi_R(0^+)=
\begin{pmatrix} 
1 & \hat{\mathbf{q}}_R^\top(\eta) \\ 
0 & \mathbb{I}_N 
\end{pmatrix},
\end{equation}
and hence
\begin{equation}
\Phi_R^{-1}(0^+)=
\begin{pmatrix} 
1 & -\hat{\mathbf{q}}_R^\top(\eta) \\ 
0 & \mathbb{I}_N 
\end{pmatrix}.
\end{equation}

\medskip
\noindent
Putting everything together, one finds
\begin{equation}
S(t=0,\eta)=
\begin{pmatrix} 
1 + \hat{\mathbf{q}}_R^\top \overline{\mathbf{r}} 
& -\hat{\mathbf{q}}_L^\top - \hat{\mathbf{q}}_R^\top -(\hat{\mathbf{q}}^\top_R\overline{\mathbf{r}})\hat{\mathbf{q}}_L^\top \\ 
-\overline{\mathbf{r}} 
& \mathbb{I}_N + \overline{\mathbf{r}}\,\hat{\mathbf{q}}_L^\top 
\end{pmatrix}.
\end{equation}

\medskip
\noindent
We now turn to the case $t=T$. The roles of $\mathbf{q}$ and $\mathbf{r}$ are exchanged, and the spatial Lax matrix becomes
\begin{equation}
\tilde{U}(x,T;\eta)=
\begin{pmatrix} 
i\eta & -\overline{\mathbf{q}}^\top \delta(x) \\ 
-\mathbf{r}(x,t=T) & -i\eta \mathbb{I}_{N} 
\end{pmatrix}.
\end{equation}

\medskip
\noindent
Proceeding exactly as before, we construct the Jost solutions. For $x<0$,
\begin{equation}
\Phi_L(x)= e^{i\Sigma x}
\begin{pmatrix} 
1 & 0 \\ 
-\displaystyle\int_{-\infty}^x e^{2i\eta y}\mathbf{r}(y,t=T)\,dy & \mathbb{I}_N 
\end{pmatrix},
\end{equation}
while for $x>0$,
\begin{equation}
\Phi_R(x)= e^{i\Sigma x}
\begin{pmatrix} 
1 & 0 \\ 
\displaystyle\int_x^{\infty} e^{2i\eta y}\mathbf{r}(y,t=T)\,dy & \mathbb{I}_N 
\end{pmatrix}.
\end{equation}

\medskip
\noindent
The discontinuity at the origin is now encoded in the jump matrix
\begin{equation}
\mathbf{J}=
\begin{pmatrix} 
1 & -\overline{\mathbf{q}}^\top \\ 
0 & \mathbb{I}_N 
\end{pmatrix}.
\end{equation}

\medskip
\noindent
Introducing
\begin{equation}
\hat{\mathbf{r}}_L(\eta)=\int_{-\infty}^0 e^{2i\eta y}\mathbf{r}(y,t=T)\,dy,\qquad
\hat{\mathbf{r}}_R(\eta)=\int_{0}^{\infty} e^{2i\eta y}\mathbf{r}(y,t=T)\,dy,
\end{equation}
and $\hat{\mathbf{r}}=\hat{\mathbf{r}}_L+\hat{\mathbf{r}}_R$, one obtains after the same matching procedure
\begin{equation}
S(t=T,\eta)=
\begin{pmatrix} 
1 + \overline{\mathbf{q}}^\top \hat{\mathbf{r}}_L  
& -\overline{\mathbf{q}}^\top\\ 
-\hat{\mathbf{r}}_L - \hat{\mathbf{r}}_R -(\overline{\mathbf{q}}^\top\hat{\mathbf{r}}_L)\hat{\mathbf{r}}_R  
& \mathbb{I}_N + \hat{\mathbf{r}}_R \overline{\mathbf{q}}^\top 
\end{pmatrix}.
\end{equation}

\section{Alternative determination of $\omega$}

We provide here an alternative computation of the parameter $\omega$.
For $\eta=0$ the potential matrix $\tilde{U}$ is pure gauge
$
\tilde{U}(x,t;\eta=0)=-A(x,t)=-G^{-1}\partial_x G
$, so the proapagator is given by 
\begin{equation}
\Phi(x,y,t) =  G^{-1}(x,t)G(y,t).
\end{equation} 
Moreover since $\lim_{x\to\pm \infty}\tilde{U}(x,t;\eta=0)=0 $ we have a scattering matrix given by
\begin{equation}
S(\eta=0)=G^{-1}_RG_L
\end{equation}
where $G_R= \lim_{x\to +\infty} G(x,t), G_L= \lim_{x\to -\infty} G(x,t)$.
Now notice that we have
\begin{equation}
\sigma_R=\lim_{x\to+\infty}\sigma(x)= G_R \Sigma G^{-1}_R,\qquad 
\sigma_L=\lim_{x\to-\infty}\sigma(x)= G_L \Sigma G^{-1}_L
\end{equation}
which are completely determined by the boundary conditions
\begin{equation}
\sigma_R = 2 \mathbf{Q}_R \mathbf{P}^\top_R-\mathbb{I}_{N+1},\qquad
\sigma_L = 2 \mathbf{Q}_L \mathbf{P}^\top_L-\mathbb{I}_{N+1}.
\end{equation}
So we have a trick to relate $\omega$ to the boundary conditions
\begin{equation}
\tr [\sigma_R\sigma_L]= 
\tr[G_R \Sigma G^{-1}_RG_L \Sigma G^{-1}_L]= 
\tr[\Sigma S(0)\Sigma S^{-1}(0)]
\end{equation}
Now
\begin{equation}
\begin{split}
\tr [\sigma_R\sigma_L]&= 4 (\mathbf{P}^\top_R\mathbf{Q}_L)(\mathbf{P}^\top_L\mathbf{Q}_R)-3+N\\
\tr[\Sigma S(0)\Sigma S^{-1}(0)]&=4\omega+N+1 
\end{split}
\end{equation}
from which we conclude
\begin{equation}
\omega= (\mathbf{P}^\top_R\mathbf{Q}_L)(\mathbf{P}^\top_L\mathbf{Q}_R)-1
=
\left(\sum_{\alpha=0}^{N}\rho_{L,\alpha}e^{\lambda_\alpha}\right)
\left(\sum_{\alpha=0}^{N}\rho_{R,\alpha}e^{-\lambda_\alpha}\right)-1.
\end{equation}

\typeout{}
\bibliographystyle{unsrt}
\bibliography{ssep-bib}

\end{document}